\documentclass{emulateapj}

\usepackage{natbib}
\usepackage{apjfonts}
\begin{document}

\title{The Lick AGN Monitoring Project: Photometric Light Curves and
Optical Variability Characteristics}

\author{Jonelle L. Walsh\altaffilmark{1}, Takeo
Minezaki\altaffilmark{2}, Misty C. Bentz\altaffilmark{1,3}, Aaron
J. Barth\altaffilmark{1}, Nairn Baliber \altaffilmark{4,5}, Weidong
Li\altaffilmark{6}, Daniel Stern\altaffilmark{7}, Vardha Nicola
Bennert\altaffilmark{4}, Timothy M. Brown\altaffilmark{5}, Gabriela
Canalizo\altaffilmark{8,9}, Alexei V. Filippenko\altaffilmark{6},
Elinor L. Gates\altaffilmark{10}, Jenny E. Greene\altaffilmark{3,11},
Matthew A. Malkan\altaffilmark{12}, Yu Sakata\altaffilmark{2,13},
Rachel A. Street\altaffilmark{4,5}, Tommaso Treu\altaffilmark{4,14},
Jong-Hak Woo\altaffilmark{3,12}, and Yuzuru Yoshii\altaffilmark{2,15}}

\altaffiltext{1}{Department of Physics and Astronomy, University of
California at Irvine, 4129 Frederick Reines Hall, Irvine, CA
92697-4574; jlwalsh@uci.edu.}

\altaffiltext{2}{Institute of Astronomy, School of Science, University
of Tokyo, 2-21-1 Osawa, Mitaka, Tokyo 181-0015, Japan.}

\altaffiltext{3}{Hubble Fellow.}

\altaffiltext{4}{Physics Department, University of California, Santa
Barbara, CA 93106.}

\altaffiltext{5}{Las Cumbres Observatory Global Telescope, 6740
Cortona Dr. Ste. 102, Goleta, CA 93117.}

\altaffiltext{6}{Department of Astronomy, University of California,
Berkeley, CA 94720-3411.}

\altaffiltext{7}{Jet Propulsion Laboratory, California Institute of
Technology, MS 169-527, 4800 Oak Grove Drive, Pasadena, CA 91109.}

\altaffiltext{8}{Department of Physics and Astronomy, University of
California, Riverside, CA 92521.}

\altaffiltext{9}{Institute of Geophysics and Planetary Physics,
University of California, Riverside, CA 92521.}

\altaffiltext{10}{Lick Observatory, P.O. Box 85, Mount Hamilton, CA
95140.}

\altaffiltext{11}{Princeton University Observatory, Princeton, NJ
08544.}

\altaffiltext{12}{Department of Physics and Astronomy, University of
California, Los Angeles, CA 90024.}

\altaffiltext{13}{Department of Astronomy, School of Science,
University of Tokyo, 7-3-1 Hongo, Bunkyo-ku, Tokyo 113-0033, Japan.}

\altaffiltext{14}{Sloan Fellow, Packard Fellow.}

\altaffiltext{15}{Research Center for the Early Universe, School of
Science, University of Tokyo, 7-3-1 Hongo, Bunkyo-ku, Tokyo 113-0033,
Japan.}

\begin{abstract}

The Lick AGN Monitoring Project targeted 13 nearby Seyfert 1 galaxies
with the intent of measuring the masses of their central black holes
using reverberation mapping. The sample includes 12 galaxies selected
to have black holes with masses roughly in the range $10^6$--$10^7$
M$_\odot$, as well as the well-studied AGN NGC 5548. In conjunction
with a spectroscopic monitoring campaign, we obtained broad-band $B$
and $V$ images on most nights from 2008 February through 2008 May. The
imaging observations were carried out by four telescopes: the 0.76-m
Katzman Automatic Imaging Telescope (KAIT), the 2-m Multicolor Active
Galactic Nuclei Monitoring (MAGNUM) telescope, the Palomar 60-in
(1.5-m) telescope, and the 0.80-m Tenagra II telescope. Having
well-sampled light curves over the course of a few months is useful
for obtaining the broad-line reverberation lag and black hole mass,
and also allows us to examine the characteristics of the continuum
variability. In this paper, we discuss the observational methods and
the photometric measurements, and present the AGN continuum light
curves. We measure various variability characteristics of each of the
light curves. We do not detect any evidence for a time lag between the
$B$- and $V$-band variations, and we do not find significant color
variations for the AGNs in our sample.

\end{abstract}


\section{Introduction}
\label{sec:intro}

The variability of active galactic nuclei (AGNs) has been studied
extensively across a wide range of wavelengths extending from radio to
X-rays. Many studies have focused on establishing simple relationships
between variability and various physical parameters. Variability
amplitude has been found to correlate with redshift $z$
\citep[e.g.,][]{Cristiani_Vio_Andreani_1990, Hook_1994, Trevese_1994,
VandenBerk_2004} and black hole mass on longer timescales of >100 days
\citep{Wold_2007, Wilhite_2008, Bauer_2009}, and anticorrelations have
been seen with luminosity \citep[e.g.,][]{Cristiani_1997, Giveon_1999,
VandenBerk_2004, Bauer_2009}, Eddington ratio \citep{Wilhite_2008,
Bauer_2009}, and rest-frame wavelength \citep[e.g.,][]{Cutri_1985,
diClemente_1996, Giveon_1999, VandenBerk_2004}.

A number of models have been developed in order to explain the
physical mechanisms driving the variability. Processes that are
external to the accretion disk have been discussed, such as
variability due to multiple supernovae or starbursts which occur close
to the nucleus \citep{Terlevich_1992, Aretxaga_Terlevich_1994},
stellar collisions \citep{TorricelliCiamponi_2000}, or gravitational
microlensing by small compact objects along the line of sight
\citep{Hawkins_1993}. Recently, several studies have shown that
optical variations on long timescales (i.e., > 100 days) are probably
the result of changes in the accretion rate \citep{Wold_2007,
Li_Cao_2008, Arevalo_2008, Wilhite_2008}. However, fluctuations in the
accretion rate cannot account for the optical variations seen on short
timescales, of order days. In order to describe optical variations on
short timescales, some have suggested that X-ray reprocessing plays a
role. In this scenario, the X-ray continuum heats the accretion disk
and changes in this X-ray continuum cause optical continuum
variability (e.g., \citealt{Krolik_1991, Sergeev_2005, Cackett_2007,
Arevalo_2008}).

Despite large observational and theoretical efforts to study
variability, the physical mechanisms behind AGN variability are still
poorly understood. However, studying AGN variability can provide
constraints on the sizes of different regions associated with the
central engine. Variability has been used very successfully in
determining the sizes of the broad-line region (BLR), and consequently
the mass of black holes in nearby Type 1 AGNs, through the technique
known as reverberation mapping \citep{Blandford_McKee_1982,
Peterson_1993}. The method uses the time delay between variations in
the AGN continuum and the subsequent response of the broad emission
lines to measure the size of the BLR. By combining the BLR size with
the velocity width of an emission line, typically the broad H$\beta$
line, the black hole mass can be determined using a simple viral
equation.

The results from reverberation mapping have led to mass measurements
of about three dozen black holes at the centers of nearby active
galaxies (\citealt{Peterson_2004}, 2005), as well as the discovery of
a correlation between the BLR size and the AGN continuum luminosity
(the $R-L$ relation; \citealt{Koratkar_Gaskell_1991,Kaspi_2000}, 2005;
\citealt{Bentz_2006}, 2009a). The $R-L$ relationship, however, is
poorly constrained at the low-luminosity end, for continuum
luminosities $\lambda L_{\lambda}(5100$ \AA $) \lesssim 10^{43}$ erg
s$^{-1}$. Additionally, the low-mass end of the relationship between
AGN black hole mass and bulge stellar velocity dispersion
($M_\mathrm{BH}-\sigma_{\star}$) is even more poorly determined since
most of the AGNs making up the relationship contain black holes with
masses between $10^7$ and $3 \times 10^8$ M$_\odot$
\citep{Onken_2004}.

In order to increase the number of reverberation measurements at the
low ends of the AGN luminosity and mass regimes, we began the Lick AGN
Monitoring Project, a reverberation-mapping campaign that targeted 12
nearby Seyfert 1 galaxies containing black holes with expected masses
between $10^6$ and $3 \times 10^7$ M$_\odot$, and also the
well-studied AGN NGC 5548 with a black hole mass of $6.54 \times 10^7$
M$_\odot$ \citep{Bentz_2007}. The project consisted of a mostly
contiguous 64-night spectroscopic monitoring campaign at the Lick
Observatory 3-m Shane telescope, along with simultaneous nightly $B$
and $V$ imaging carried out at four smaller telescopes. Most previous
reverberation campaigns measured the AGN continuum variations directly
from the spectra. However, for our sample of relatively low-luminosity
AGNs, measuring the continuum flux from broad-band images is more
accurate, as we are able to obtain a better calibration and a higher
signal-to-noise ratio using images rather than spectra.

Having regularly sampled light curves with a daily cadence over the
course of a few months is useful for determining broad-line
reverberation time lags and black hole masses, and also allows us to
examine optical continuum variability characteristics. In this paper,
we present the results from the imaging campaign. We discuss the
sample selection, describe the broad-band observations, and provide
details of the aperture photometry. We present the $B$- and $V$-band
continuum light curves for the 13 objects, cross-correlate them to
search for time delays between the two bands, and attempt to
investigate the $B-V$ color for the AGNs. In companion papers, these
AGN continuum light curves are compared to the light curves of
H$\beta$ \citep[][hereafter Paper III]{Bentz_2009b} and other optical
recombination lines \citep{Bentz_2009c} to measure BLR sizes and black
hole masses. The first results from the reverberation mapping project
on the object Arp 151 were previously presented \citep[][hereafter
Paper I]{Bentz_2008}.

\section{Sample Selection}
\label{sec:sample}

The main goal of the Lick AGN Monitoring Project is to substantially
increase the number of reverberation-based black hole masses at the
low end of the AGN mass and luminosity range. We selected a sample of
12 nearby ($z < 0.05$) Seyfert 1 galaxies with a single-epoch mass
estimate of $10^6 < M_\mathrm{BH} < 3 \times 10^7$ M$_\odot$ based on
the broad H$\beta$ line width. The objects were selected from
\cite{Greene_Ho_2007} and from our own library of AGN spectra obtained
at the Palomar, Lick, and Keck observatories. The estimated time lag
based on the 5100~\AA\ continuum luminosity for these AGNs is 3--20
days, which allowed for a monitoring campaign of manageable length,
extending over a few months with a nightly cadence. The objects were
also required to have a strong, broad H$\beta$ emission line suitable
for spectroscopic monitoring. In addition to these 12 objects, NGC
5548, which is among the best-studied reverberation objects and known
to be highly variable with a well-determined black hole mass
\citep[and references therein]{Bentz_2007}, was also included in our
sample as a control object. By adding NGC 5548 to our sample, we are
able to expand the existing set of data on NGC 5548 as well as more
reliably compare our new results to those from previous
reverberation-mapping campaigns. Properties of all 13 AGNs in the
sample are given in Table \ref{tab:galprop}.


\begin{deluxetable*}{lccccc}
\tabletypesize{\scriptsize} 
\tablewidth{0pt} 
\tablecaption{Properties of AGNs in the Sample \label{tab:galprop}} 
\tablehead{
\colhead{Object} & 
\colhead{$\alpha$ (J2000.0)} & 
\colhead{$\delta$ (J2000.0)} & 
\colhead{$z$} & 
\colhead{$A_\mathrm{B}$} &
\colhead{Alternate}\\
\colhead{} & 
\colhead{} & 
\colhead{} & 
\colhead{} & 
\colhead{(mag)} &
\colhead{Name}
}

\startdata

Mrk 142       & 10 25 31.3 & +51 40 35 & 0.04494 & 0.069 & PG 1022+519 \\
SBS 1116+583A & 11 18 57.7 & +58 03 24 & 0.02787 & 0.050 & \\
Arp 151       & 11 25 36.2 & +54 22 57 & 0.02109 & 0.059 & Mrk 40 \\
Mrk 1310      & 12 01 14.3 & -03 40 41 & 0.01941 & 0.133 & \\
Mrk 202       & 12 17 55.0 & +58 39 35 & 0.02102 & 0.087 & \\
NGC 4253      & 12 18 26.5 & +29 48 46 & 0.01293 & 0.084 & Mrk 766 \\
NGC 4748      & 12 52 12.4 & -13 24 53 & 0.01463 & 0.223 & \\
IC 4218       & 13 17 03.4 & -02 15 41 & 0.01933 & 0.132 & \\
MCG-06-30-15  & 13 35 53.8 & -34 17 44 & 0.00775 & 0.266 & ESO 383-G035 \\
NGC 5548      & 14 17 59.5 & +25 08 12 & 0.01718 & 0.088 & \\
Mrk 290       & 15 35 52.3 & +57 54 09 & 0.02958 & 0.065 & PG 1534+580 \\
IC 1198       & 16 08 36.4 & +12 19 51 & 0.03366 & 0.236 & Mrk 871 \\
NGC 6814      & 19 42 40.6 & -10 19 25 & 0.00521 & 0.790 & \\

\enddata

\tablecomments{Units of right ascension are hours, minutes, seconds,
and units of declination are degrees, arcminutes,
arcseconds. Redshifts come the NASA Extragalactic Database (NED). The
Galactic extinction, $A_\mathrm{B}$, is given by
\cite{Schlegel_1998}.}

\end{deluxetable*}

\section{Observations}
\label{sec:obs}

Broad-band Johnson $B$ and $V$ images were obtained about twice a week
from 2008 February 9 to 2008 March 15, and then on most nights between
2008 March 16 and 2008 June 1 (UT dates are used throughout this
paper). We began the photometric monitoring prior to the onset of the
spectroscopic campaign (which began on 2008 March 25) since the
response of the broad-line emission is delayed relative to the
variations in the continuum flux. In Table \ref{tab:ltcurvestats}, we
give typical exposure times and statistical properties of the light
curves for each object. The sample was divided between four
telescopes, such that each site was primarily responsible for
observing a subset of the objects. In Table \ref{table:telprop}, we
list the properties of the four telescopes: the Katzman Automatic
Imaging Telescope (KAIT), the Multicolor Active Galactic Nuclei
Monitoring (MAGNUM) telescope, the Palomar 60-in telescope (P60), and
the Tenagra II telescope. Additionally, standard stars for the flux
calibration of several light curves were observed at the Faulkes
Telescope North (FTN) on three photometric nights between 2008 June 25
and 2008 July 1.


\begin{deluxetable*}{lccccccccccc}
\tabletypesize{\scriptsize} 
\tablewidth{0pt} 
\tablecaption{Light-Curve Statistics \label{tab:ltcurvestats}} 
\tablehead{
\colhead{} &
\multicolumn{5}{c}{$B$ Band} &&
\multicolumn{5}{c}{$V$ Band} \\
\cline{2-6}\cline{8-12} \\
\colhead{Object} & 
\colhead{Exposure Time} &
\colhead{$N$} & 
\colhead{$T_\mathrm{median}$} & 
\colhead{UT Start} & 
\colhead{UT End} &
\colhead{} & 
\colhead{Exposure Time} &
\colhead{$N$} &
\colhead{$T_\mathrm{median}$} &
\colhead{UT Start} &
\colhead{UT End} \\
\colhead{} &
\colhead{(s)} &
\colhead{} &
\colhead{(days)} &
\colhead{} &
\colhead{} &
\colhead{} &
\colhead{(s)} &
\colhead{} &
\colhead{(days)} &
\colhead{} &
\colhead{} \\
\colhead{(1)} &
\colhead{(2)} &
\colhead{(3)} &
\colhead{(4)} &
\colhead{(5)} &
\colhead{(6)} &
\colhead{} &
\colhead{(7)} &
\colhead{(8)} &
\colhead{(9)} &
\colhead{(10)} &
\colhead{(11)}
}

\startdata

Mrk 142       & 2$\times$300 & 80 & 0.98 & 2008-02-10 & 2008-06-01 && 2$\times$300 & 76 & 0.98 & 2008-02-10 & 2008-05-24 \\
SBS 1116+583A & 2$\times$300 & 56 & 1.02 & 2008-02-09 & 2008-06-01 && 2$\times$200 & 56 & 1.01 & 2008-02-09 & 2008-05-24 \\
Arp 151       & 2$\times$300 & 84 & 0.93 & 2008-02-10 & 2008-05-16 && 2$\times$300 & 78 & 0.96 & 2008-02-10 & 2008-05-16 \\
Mrk 1310      & 4$\times$130 & 50 & 1.16 & 2008-02-20 & 2008-05-30 && 4$\times$130 & 58 & 1.05 & 2008-02-20 & 2008-05-30 \\
Mrk 202       & 450 & 58 & 1.01 & 2008-02-09 & 2008-06-01 && 200 & 58 & 1.01 & 2008-02-09 & 2008-05-24 \\
NGC 4253      & 2$\times$300 & 58 & 1.00 & 2008-02-13 & 2008-05-16 && 2$\times$300 & 63 & 0.97 & 2008-02-13 & 2008-05-16 \\
NGC 4748      & 2$\times$120 & 48 & 1.12 & 2008-02-09 & 2008-06-02 && 2$\times$90 & 52 & 1.03 & 2008-02-09 & 2008-06-01 \\
IC 4218       & 2$\times$120 & 42 & 1.03 & 2008-02-09 & 2008-06-01 && 2$\times$120 & 56 & 1.09 & 2008-02-09 & 2008-06-01 \\
MCG-06-30-15  & 4$\times$45 & 48 & 1.08 & 2008-02-20 & 2008-05-30 && 4$\times$30 & 55 & 1.04 & 2008-02-20 & 2008-05-30 \\ 
NGC 5548      & 6$\times$50 & 45 & 1.08 & 2008-02-12 & 2008-05-29 && 6$\times$35 & 57 & 1.05 & 2008-02-12 & 2008-05-30 \\
Mrk 290       & 300 & 50 & 1.01 & 2008-02-09 & 2008-06-01 && 150 & 50 & 1.01 & 2008-02-09 & 2008-05-23 \\                
IC 1198       & 2$\times$300 & 61 & 1.02 & 2008-02-13 & 2008-06-01 && 2$\times$300 & 66 & 1.02 & 2008-02-13 & 2008-05-23 \\
NGC 6814      & 4$\times$45 & 43 & 1.04 & 2008-03-20 & 2008-05-31 && 4$\times$30 & 46 & 1.02 & 2008-03-20 & 2008-05-31 \\

\enddata

\tablecomments{Columns 2 and 7 list the typical exposure times for
each object. Columns 3 and 8 give the number of observations, and
columns 4 and 9 give the median sampling rate. UT dates are given in
the format YYYY-MM-DD.}

\end{deluxetable*}


\begin{deluxetable*}{lcccc}
\tabletypesize{\scriptsize} 
\tablewidth{0pt} 
\tablecaption{Telescope Properties \label{table:telprop}} 
\tablehead{
\colhead{Telescope} & 
\colhead{Mirror Diameter} & 
\colhead{Field of View} & 
\colhead{Pixel Scale} & 
\colhead{Galaxies Observed} \\
\colhead{} & 
\colhead{} & 
\colhead{} & 
\colhead{(\arcsec pixel$^{-1}$)} & 
\colhead{}
}

\startdata

KAIT & 0.8 m & 6\farcm8 $\times$ 6\farcm8 & 0.80 &
SBS $1116+583$A, Mrk 202, Mrk 290 \\

MAGNUM & 2 m & 1\farcm5 $\times$ 1\farcm5 & 0.28 &
Mrk 1310, MCG-06-30-15, NGC 5548, NGC 6814 \\

P60 & 1.5 m & 12\farcm9 $\times$ 12\farcm9 & 0.38 &
NGC 4748, IC 4218 \\

Tenagra & 0.8 m & 15\arcmin $\times$ 15\arcmin & 0.87 &
Mrk 142, Arp 151, NGC 4253, IC 1198 \\

\enddata
\end{deluxetable*}

\subsection{KAIT}
\label{subsec:kait}

KAIT is a 0.76-m robotic telescope \citep{Filippenko_2001} located at
Lick Observatory. The images were taken with a $512 \times 512$ pixel
SITe CCD. The pixel scale is 0\farcs80 pixel$^{-1}$, providing a field
of view of 6\farcm8 $\times$ 6\farcm8. During a portion of the
campaign, from 2008 April 12 to 2008 April 25, a similar camera was
used in place of the primary camera due to technical problems.

For the duration of the campaign, KAIT observed three AGNs: SBS
$1116+583$A, Mrk 202, and Mrk 290. When possible, KAIT also observed
NGC 4748, IC 4218, and IC 1198, each for $250$ s in the $V$ band, as a
backup to the P60 and Tenagra telescopes to help ensure that the AGNs
were monitored with a nightly cadence. During the last week of the
campaign, from 2008 May 27 to 2008 June 1, the P60 and Tenagra
telescopes were unable to observe their assigned objects due to
scheduling conflicts at Palomar and technical problems at Tenagra.
Consequently, during this time, KAIT observed Mrk 142, NGC 4748, IC
4218, and IC 1198 on a nightly basis in the $B$ band only, each for
$450$ s. Observations of SBS 1116, Mrk 202, and Mrk 290 were also
limited to the $B$ band during this period.

\subsection{MAGNUM}
\label{subsec:magnum}

MAGNUM is a 2-m telescope \citep{Kobayashi_1998b, Yoshii_2002} located
at Haleakala Observatory on Maui. The observations were made using the
multicolor imaging photometer \citep[MIP;][]{Kobayashi_1998a}, which
provides simultaneous optical and near-infrared images. The optical
detector is a $1024 \times 1024$ pixel SITe CCD with a scale of
0\farcs28 pixel$^{-1}$ and a field of view of 1\farcm5 $\times$
1\farcm5.

MAGNUM monitored four targets for the duration of the campaign: Mrk
1310, MCG-06-30-15, NGC 5548, and NGC 6814. Infrared images were also
simultaneously obtained for all four AGNs, and the data will be
described in a future paper. Despite MAGNUM's small field of view,
comparison stars for Mrk 1310, MCG-06-30-15, and NGC 6814 were
simultaneously observed along with the target AGN. However, the
comparison stars for NGC 5548 did not fall within the field of view,
and were instead observed with an alternating pattern as described by
\cite{Suganuma_2006}. The comparison stars for NGC 5548 were
calibrated using photometric observations of \citet{Landolt_1992}
standard stars (see \citealt{Suganuma_2006}). The flux-calibrated
comparison stars for NGC 5548 were then used to calibrate the
comparison stars for the remaining three objects.

\subsection{P60}
\label{subsec:palomar}

The Palomar 60-in (1.5-m) telescope \citep[P60;][]{Cenko_2006} is
equipped with a SITe $2048 \times 2048$ pixel CCD, with a pixel scale
of 0\farcs38 pixel$^{-1}$ and a 12\farcm9 $\times$ 12\farcm9 field of
view. Over the course of the campaign, the P60 observed two objects:
NGC 4748 and IC 4218. Photometric observations of the comparison stars
for NGC 4748 and IC 4218 were obtained by MAGNUM at the conclusion of
the campaign, and their fluxes were calibrated using the
\citet{Landolt_1992} standard star SA-105 815.

\subsection{Tenagra}
\label{subsec:tenagra}

The 0.80-m Tenagra II telescope is located at the Tenagra
Observatories complex in southern Arizona. The science camera contains
a SITe $1024 \times 1024$ pixel CCD, with a scale of 0 \farcs87
pixel$^{-1}$, yielding a field of view of 15\arcmin $\times$
15\arcmin.

Throughout the campaign, Tenagra monitored four AGNs: Mrk 142, Arp
151, NGC 4253, and IC 1198. Often, multiple epochs of data were
obtained each night. \citet{Landolt_1992} SA-101 and SA-109
standard-star fields were observed a total of 16 times at a range of
airmasses in each band on one photometric night, and these
observations were used to flux calibrate the light curves for the four
AGNs.

\subsection{FTN}
\label{subsec:ftn}

The Faulkes Telescope North is a 2-m telescope located at Haleakala
Observatory. The science camera has a field of view of 5\arcmin
$\times$ 5\arcmin\ with a pixel scale of 0\farcs28 pixel$^{-1}$. Since
KAIT did not observe standard stars for this program, after the
completion of the campaign FTN was used to observe
\citet{Landolt_1992} fields on photometric nights for the flux
calibration of the SBS 1116, Mrk 202, and Mrk 290 light curves. The
\citet{Landolt_1992} standard areas PG1323-086 and SA-107 were each
observed twice for the flux calibration of Mrk 290. On a separate
night, FTN observed the \citet{Landolt_1992} standard area SA-104 four
times at a range of airmasses for the flux calibration of Mrk
202. Finally, on a third night, SA-104 was observed twice for the flux
calibration of SBS 1116.

\section{Data Reduction and Photometry}
\label{sec:phot}

The images were reduced using the automatic pipelines from each of the
telescopes. The pipelines incorporated standard data-reduction
methods, such as overscan and bias subtraction, and flat fielding. The
P60 CCD is read out using two amplifiers; thus, an additional
data-reduction step, which combines the separate image extensions
produced by the two amplifiers into a single image, was performed
after the overscan subtraction. The P60 reduction pipeline also
included sky subtraction. As a final step, cosmic rays were removed
from the KAIT, Tenagra, and P60 images using the LA-COSMIC task
\citep{vanDokkum_2001}, and from the MAGNUM images by a manual
procedure.

\subsection{KAIT, P60, and Tenagra Photometry}
\label{subsec:othertelphot}

For each of the individual images, the flux of the AGN and the
background were measured through a circular aperture and a surrounding
annulus using the IRAF task {\tt phot} within the {\tt daophot}
package. The AGN flux was then compared to stars within the field. We
checked the consistency of the comparison stars by plotting each
star's magnitude as a function of time. Any stars that appeared to
fluctuate were not included as comparison stars. For the AGNs observed
with KAIT and P60, the AGN magnitude was determined by averaging over
the multiple exposures taken during a single night. Tenagra often
obtained several individual observations of an AGN throughout the
course of a night, and in these instances, the average AGN magnitude
was calculated for exposures separated by less than one hour.

While there exists an optimal photometric aperture size that maximizes
the signal-to-noise ratio for point sources \citep{Howell_1989}, for
extended sources the choice of aperture size is less
clear. Fluctuations in the seeing affect the extended host galaxy less
than they affect a point source, which can introduce spurious
variability in the AGN flux. Therefore, the aperture needs to be large
enough to minimize the effect of variations in the seeing, but also
small enough so that excessive amounts of noise from the sky
background and host-galaxy light are not added to the measurements.

In order to determine the ideal aperture size, we experimented with a
range of sizes for each of the objects, while keeping the same sky
annulus. We used apertures and sky annuli whose radii were an integer
number of pixels. We ultimately chose a single aperture size for each
set of AGNs observed with the same telescope. The best aperture size
was determined by selecting a representative AGN for each telescope,
cross-correlating the $B$-band photometric light curve and the
H$\beta$ emission-line light curve (see Paper III), and calculating
the time lag. The aperture size which resulted in the smallest formal
error on the time lag was selected as the best choice. The final
apertures had radii of 3\farcs20 for the three AGNs observed with
KAIT, 2\farcs65 for the two AGNs observed with P60, and 4\farcs35 for
the four AGNs observed with Tenagra.

When P60 or Tenagra were unable to observe their assigned targets due
to weather conditions, technical problems, or scheduling conflicts, we
were sometimes able to obtain the observations with KAIT. We used the
same comparison stars and the same photometric aperture size as was
used for the images from the primary telescope, but the flux measured
using the KAIT images appeared to be systematically lower than the
flux measured using the P60 images. This offset is most likely the
result of different filter transmission curves between the two
telescopes, since the transmission curves for the broad-band filters
used by P60 are known to deviate significantly from standard Johnson
transmission curves \citep{Cenko_2006}. In order to correct the issue,
we compared the flux measured with a set of KAIT and P60 images taken
on the same night. In total, there were 18 nights when both KAIT and
P60 observed NGC 4748 in the $V$ band and one night when both
telescopes observed NGC 4748 in the $B$ band. Similarly, KAIT and P60
simultaneously observed IC 4218 a total of 17 nights in the $V$ band
and three nights in the $B$ band. We determined the average offset for
the set of pairs of points and found that for NGC 4748,
$m_\mathrm{KAIT} - m_\mathrm{P60} = 0.06$ mag in the $B$ band and
0.007 mag in the $V$ band. For IC 4218, we found that $m_\mathrm{KAIT}
- m_\mathrm{P60} = 0.17$ mag in the $B$ band and 0.14 mag in the $V$
band. We adjusted the KAIT measurements accordingly. We applied a
similar comparison between a set of observations taken with KAIT and
Tenagra, but did not find evidence for a significant offset; hence, no
adjustments were made to the KAIT measurements before combining them
with the Tenagra measurements.

Finally, the $B$- and $V$-band light curves were flux calibrated. We
first determined the fluxes of the comparison stars using
\citet{Landolt_1992} standard stars, as described in \S \ref{sec:obs},
and fit the constant, color, and extinction terms of the
transformation equation over the data sets where both the comparison
stars and the \citet{Landolt_1992} standard stars were observed in the
same night. We determined the offset between the instrumental
magnitude and the calibrated magnitude for each one of the comparison
stars and then applied the average offset to the AGN light curve.

\begin{figure*}
\begin{center}
\epsscale{0.32}
\plotone{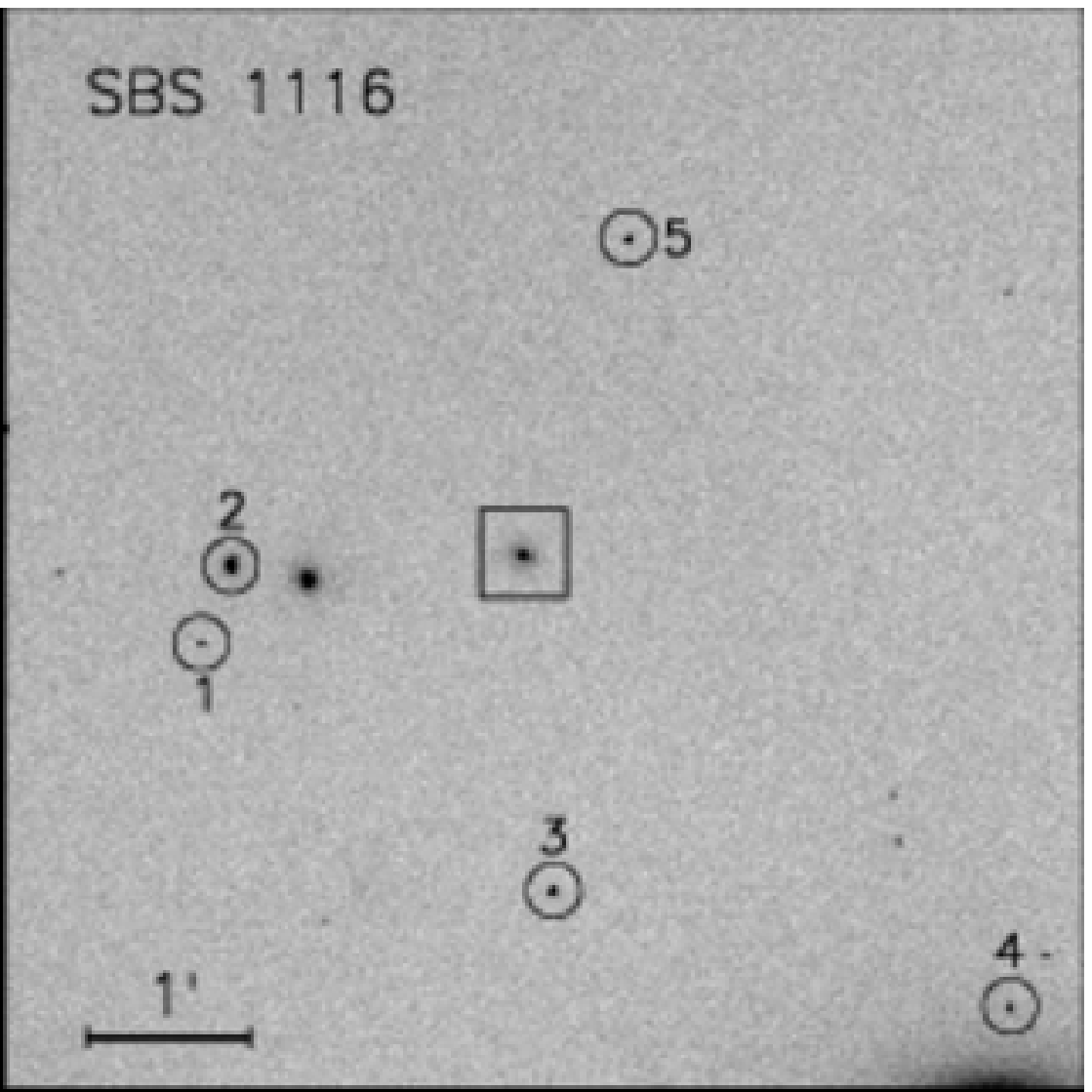}
\plotone{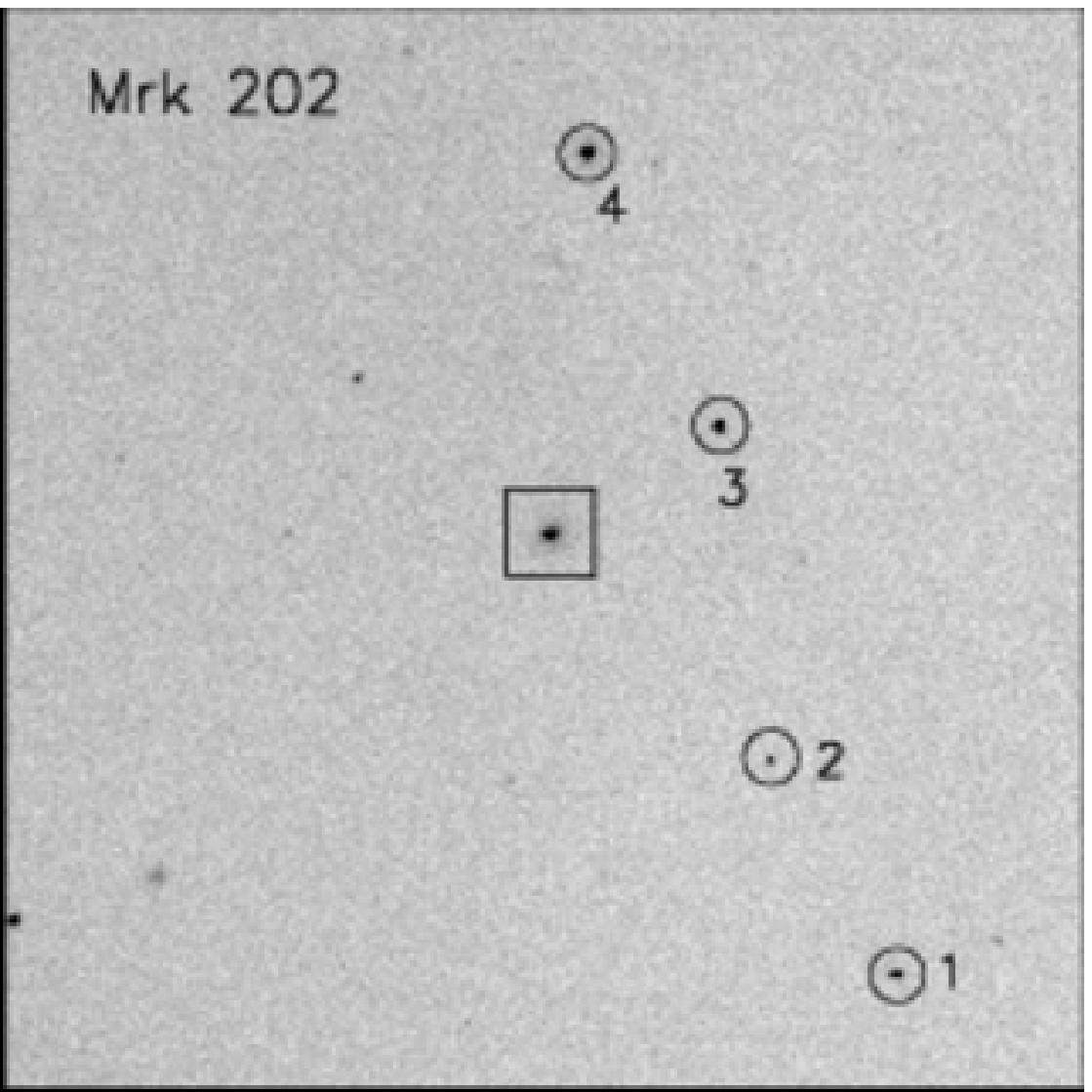}
\plotone{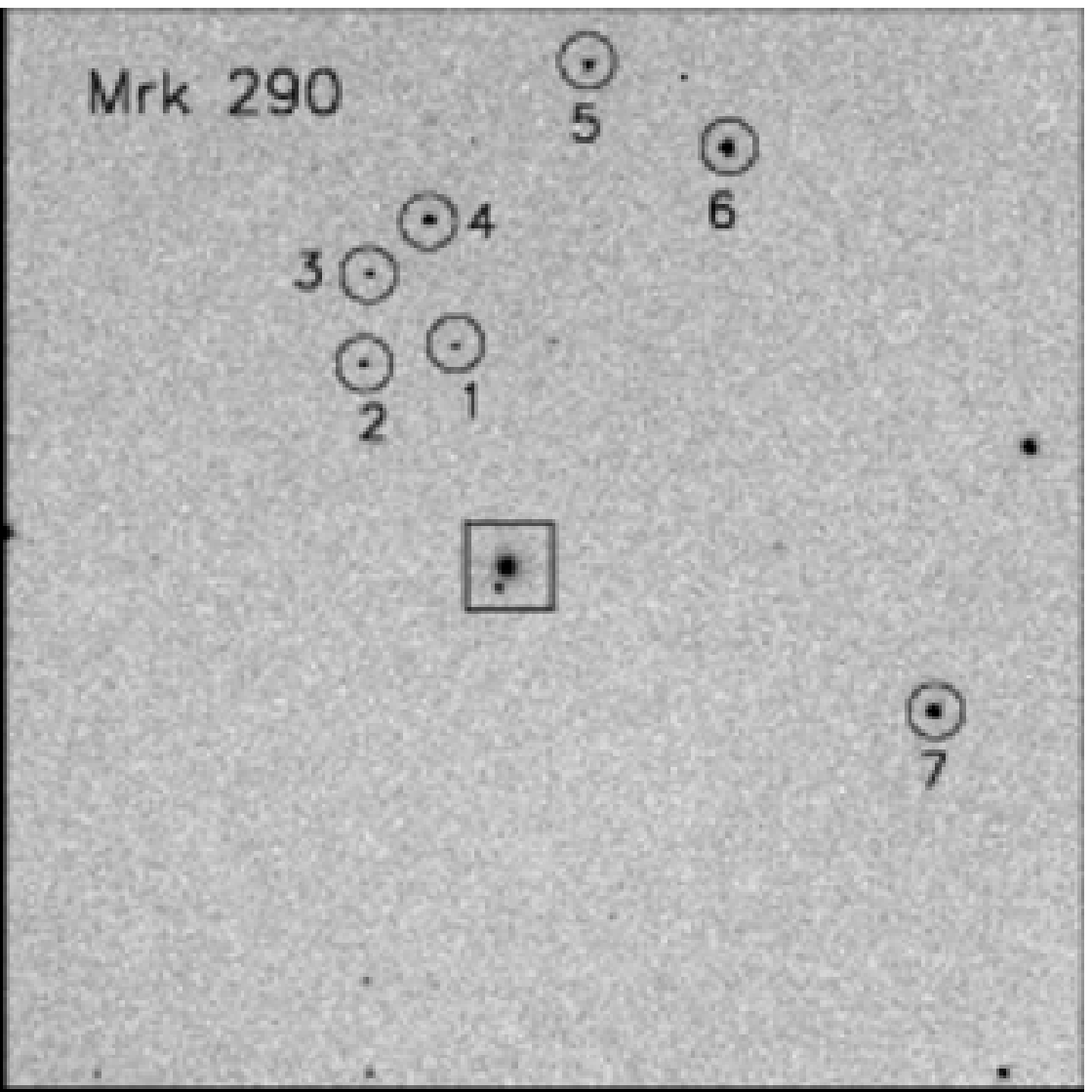}
\caption{$V$-band images of SBS 1116 (left panel), Mrk 202 (middle
panel), and Mrk 290 (right panel) taken with KAIT. North is up and
east is to the left. The scale is the same for all three images. The
square is centered on the AGN host galaxy and the numbers denote the
comparison stars. \label{fig:kaitimages}}
\end{center}
\end{figure*}

\begin{figure*}
\begin{center}
\epsscale{0.32}
\plotone{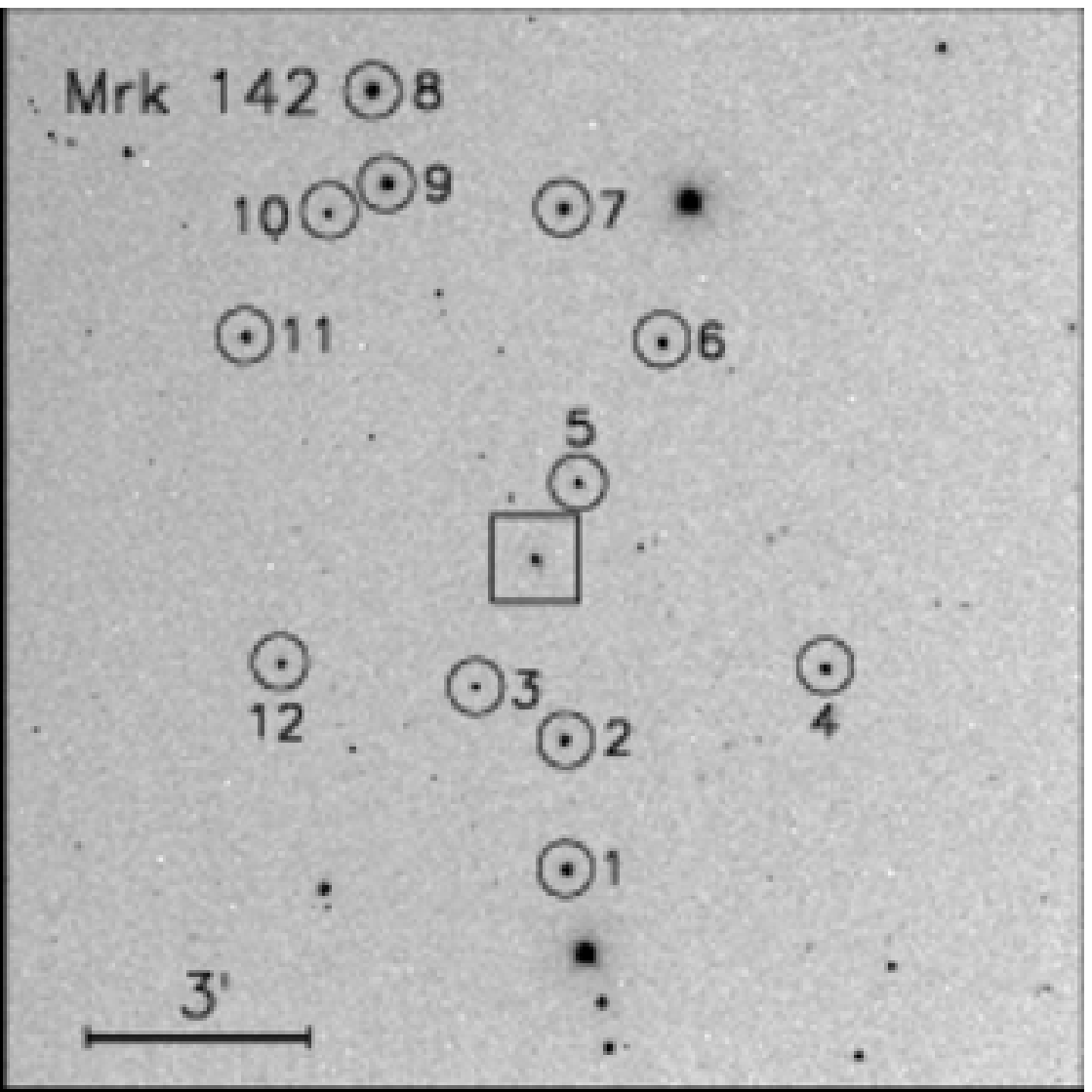}
\plotone{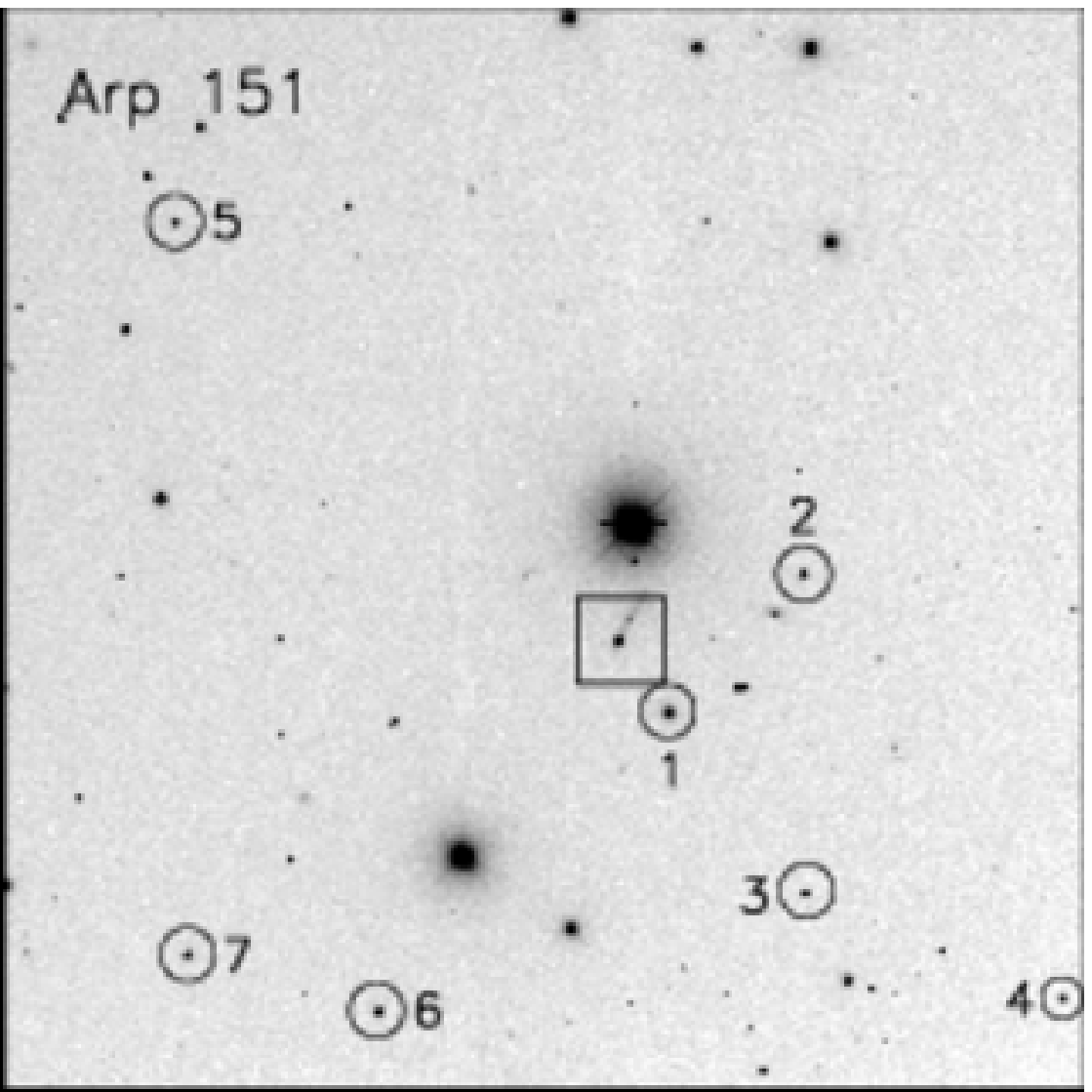} \\
\vspace{0.1cm}
\plotone{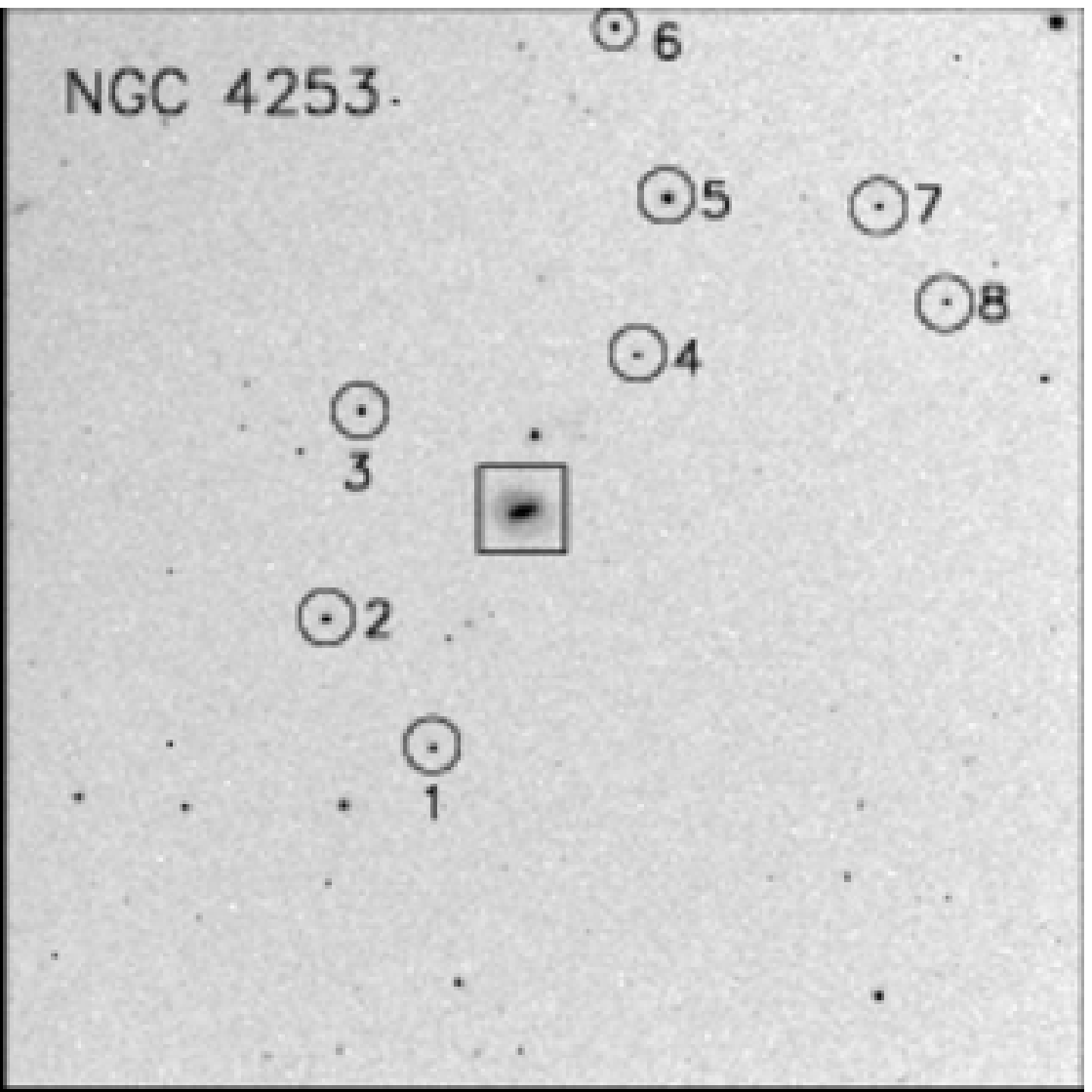}
\plotone{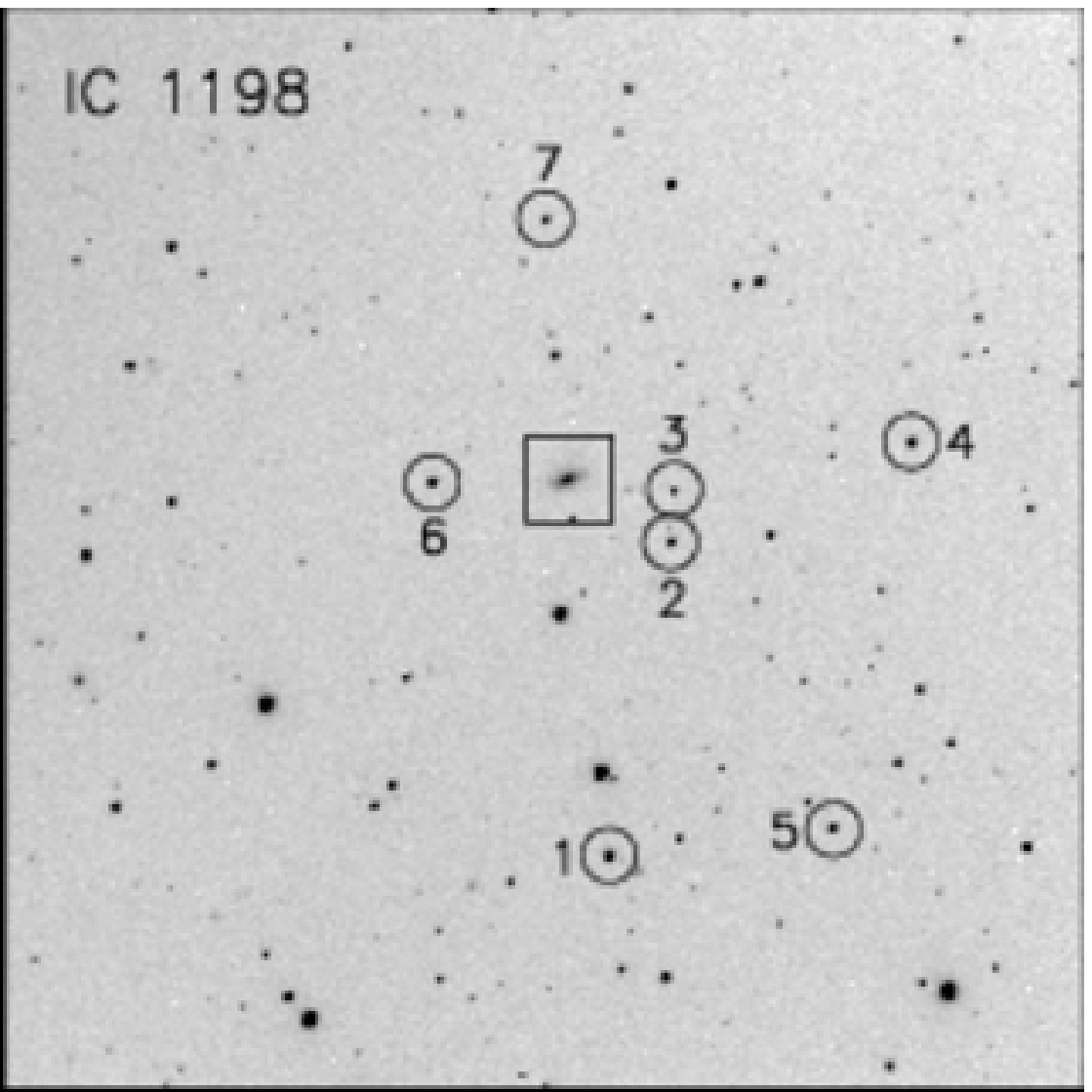}
\caption{$V$-band images of Mrk 142 (top-left panel), Arp 151 
(top-right panel), NGC 4253 (bottom-left panel), and IC 1198 (bottom-right
panel) taken with Tenagra. North is up and east is to the left. The
scale is the same for all four images. The square is centered on the
AGN host galaxy and the numbers denote the comparison
stars. \label{fig:tenagraimages}}
\end{center}
\end{figure*}

\begin{figure*}
\begin{center}
\epsscale{0.32}
\plotone{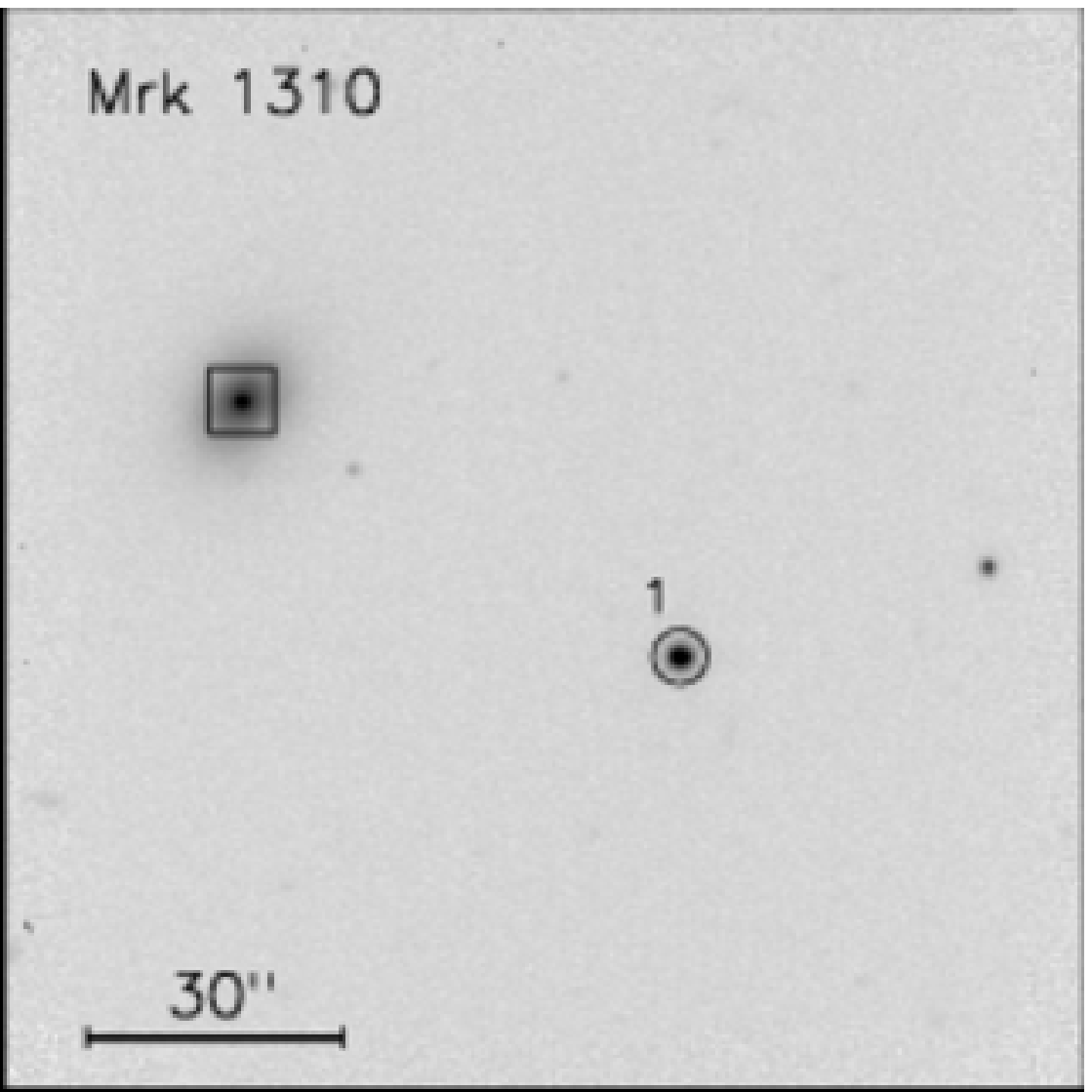}
\plotone{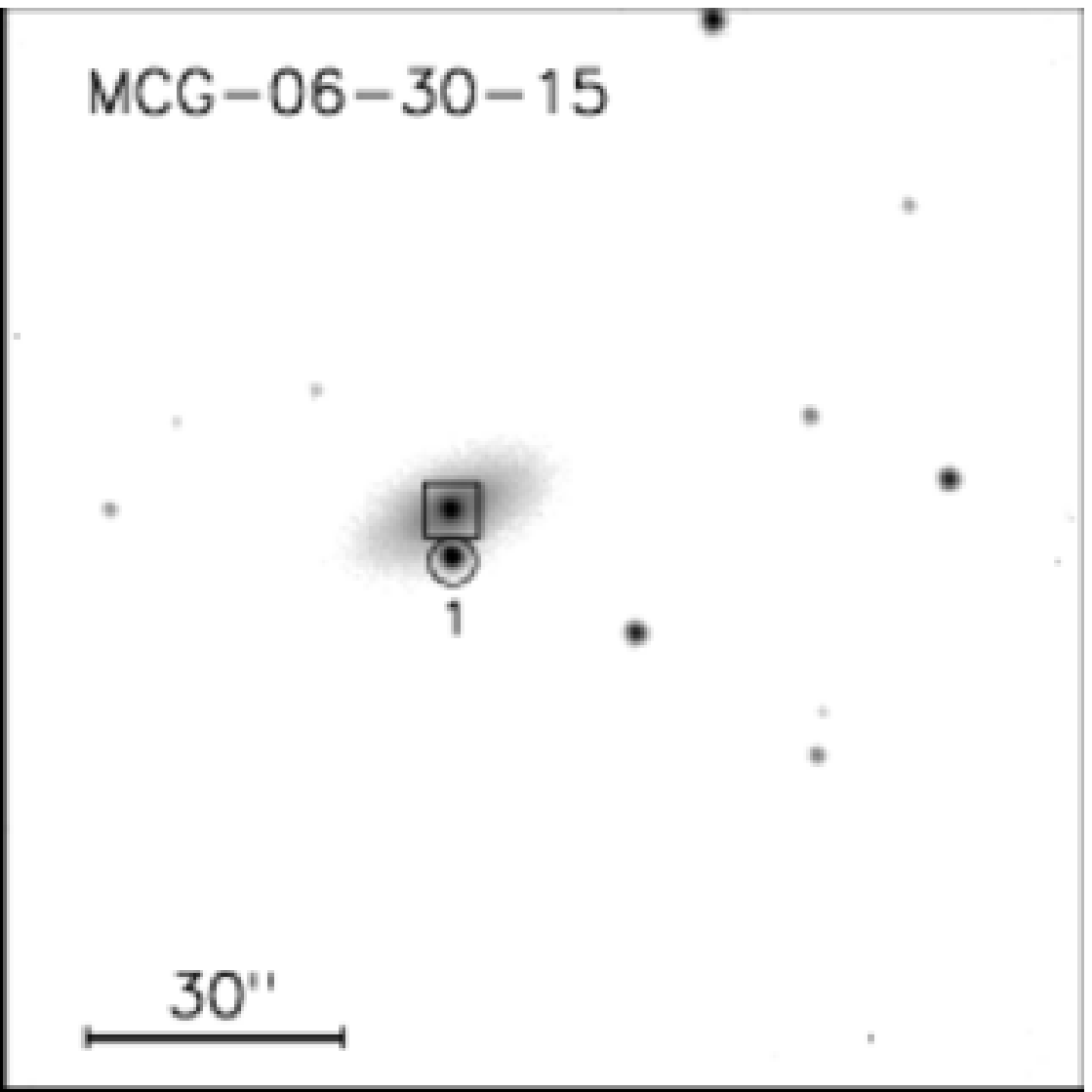} \\
\vspace{0.1cm}
\plotone{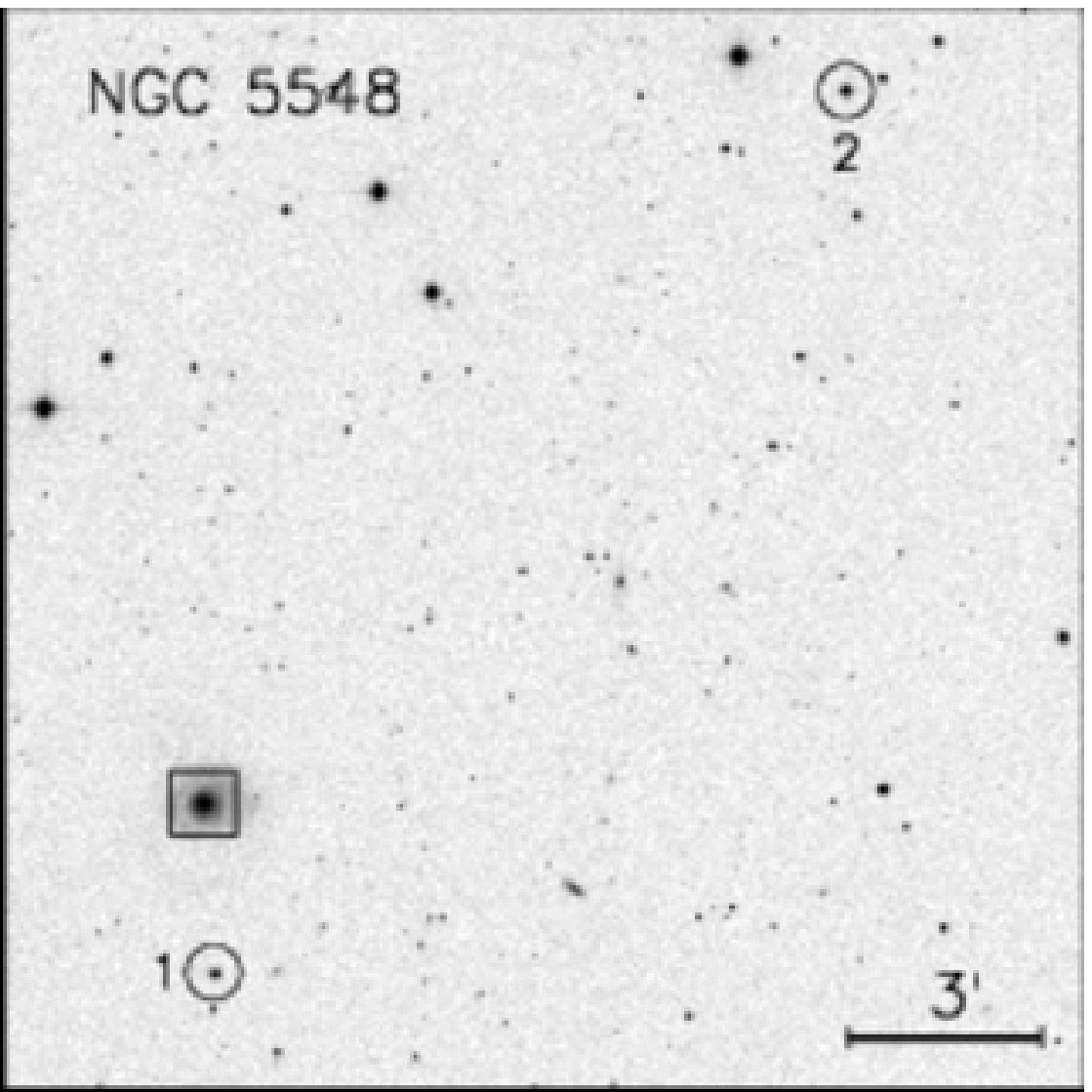}
\plotone{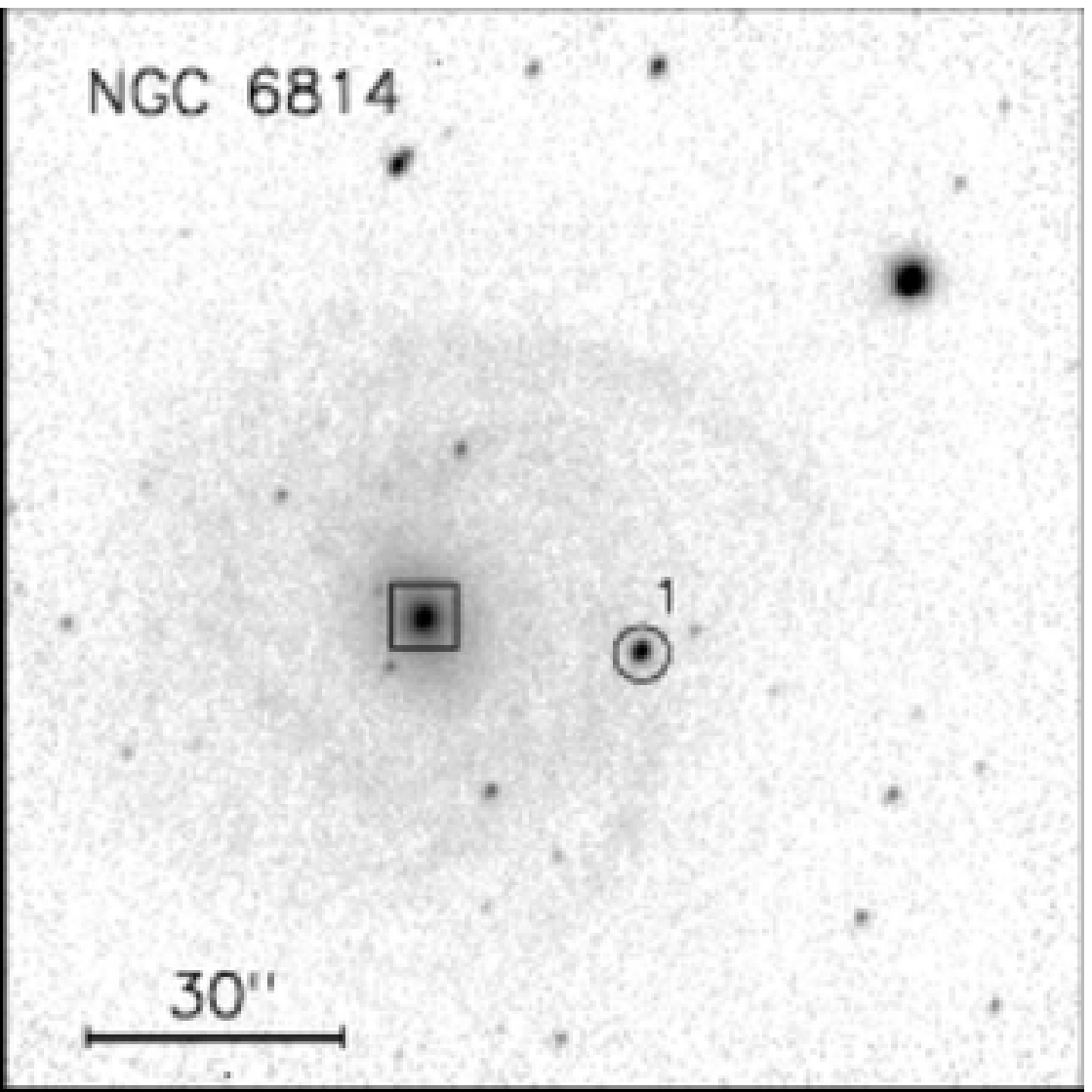}
\caption{$V$-band images of Mrk 1310 (top-left panel), MCG-06-30-15 
(top-right panel), and NGC 6814 (bottom-right panel) taken with MAGNUM.
The NGC 5548 image (bottom-left panel) was taken from the Digitized
Sky Survey (DSS). On all the images, north is up and east is to the
left. The square is centered on the AGN host galaxy and the numbers
denote the comparison stars.  For NGC 5548, the two comparison stars
do not fall within the MAGNUM field of view and were observed
alternately with the AGN.\label{fig:magnumimages}}
\end{center}
\end{figure*}

\begin{figure*}
\begin{center}
\epsscale{0.32}
\plotone{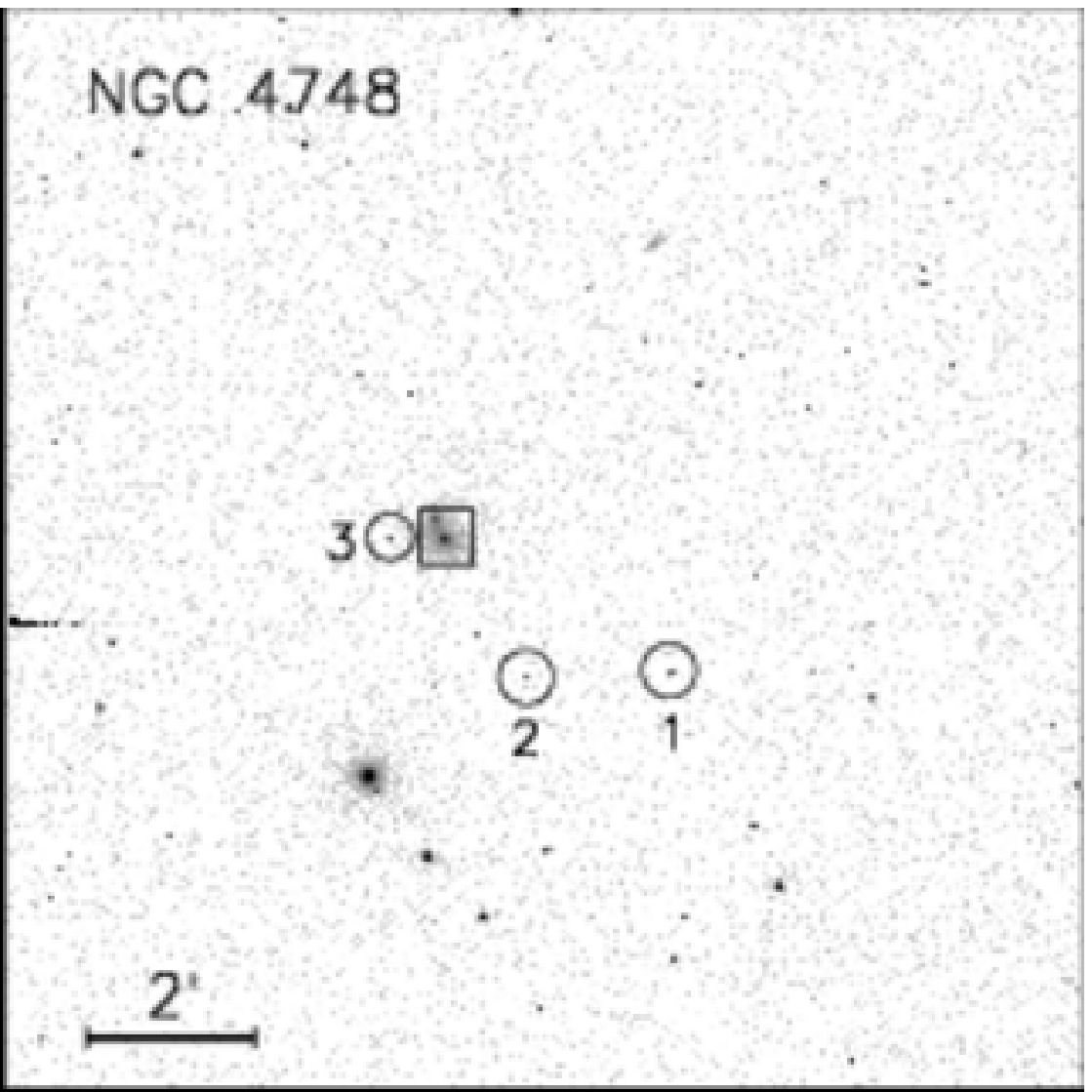}
\plotone{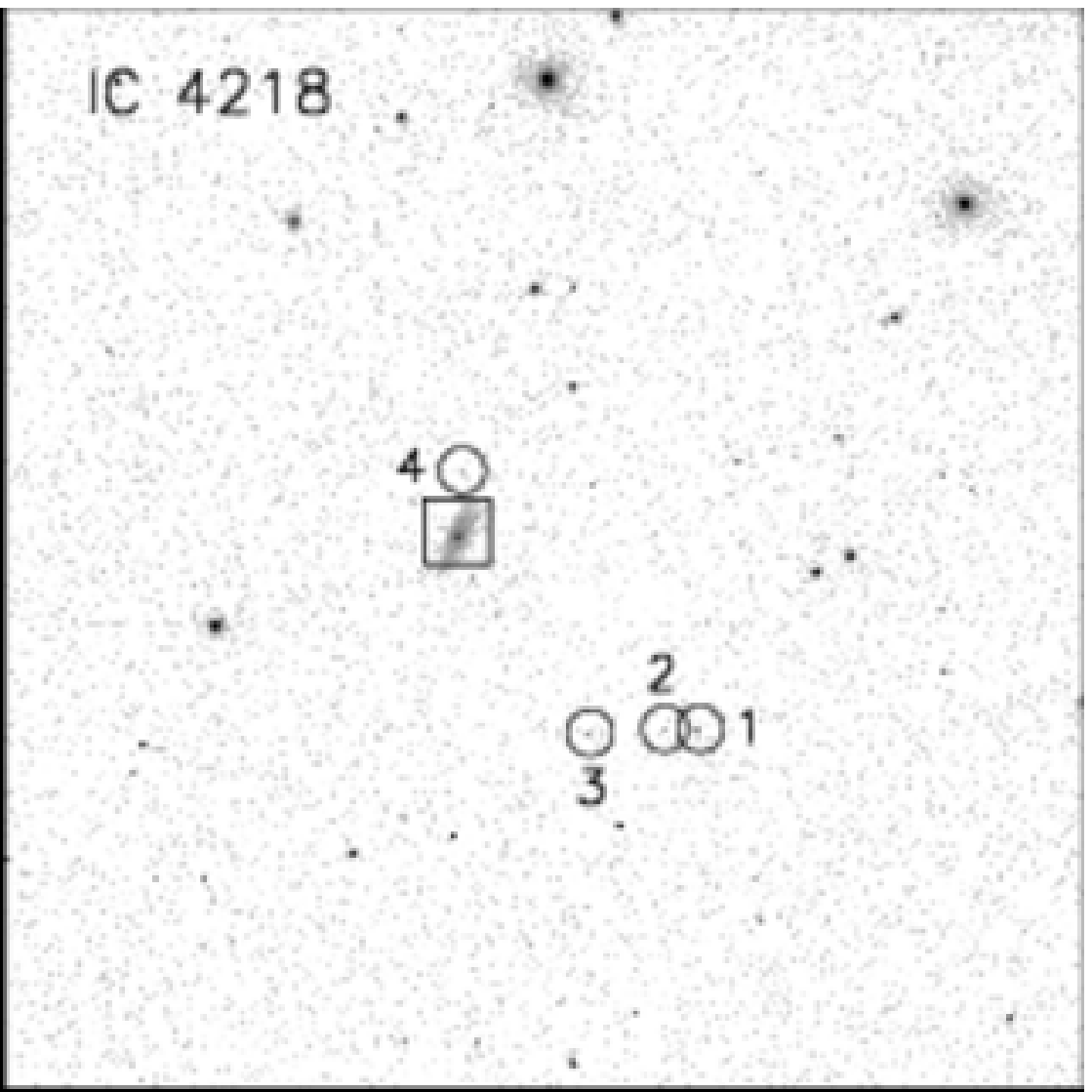}
\caption{$V$-band images of NGC 4748 (left panel) and IC 4218 (right
panel) taken with P60. North is up and east is to the left. The scale
is the same for both images. The square is centered on the AGN host
galaxy and the numbers denote the comparison
stars.\label{fig:p60images}}
\end{center}
\end{figure*}

While the calibration of the comparison stars relative to one another
and to the AGN is accurate, the absolute calibration is more
uncertain, especially in the cases when only a small number of Landolt
fields were observed over a limited airmass range from a single
night. We estimate the accuracy of the photometric zeropoint
calibration to be better than $\sim 0.08$ mag, which we determined by
measuring the standard deviation of the various measurements of a
single comparison star over the course of a photometric night. The
absolute flux calibration, however, does not affect our results since
we are only interested in relative changes. The final comparison stars
for each object are shown in Figures
\ref{fig:kaitimages}--\ref{fig:p60images} and their calibrated
magnitudes are listed in Table \ref{tab:compstars}.


\begin{deluxetable}{llcc}
\tabletypesize{\scriptsize} 
\tablewidth{0pt} 
\tablecaption{Comparison Stars \label{tab:compstars}} 
\tablehead{
\colhead{Object} & 
\colhead{Star} & 
\colhead{$B$} &
\colhead{$V$} \\ 
\colhead{} &
\colhead{} & 
\colhead{(mag)} &
\colhead{(mag)}
}
                
\startdata

Mrk 142              & 1 & 14.672 $\pm$ 0.004 & 14.023 $\pm$ 0.007  \\
                     & 2 & 15.191 $\pm$ 0.005 & 14.658 $\pm$ 0.006  \\
                     & 3 & 16.648 $\pm$ 0.053 & 16.063 $\pm$ 0.014  \\
                     & 4 & 15.303 $\pm$ 0.015 & 14.604 $\pm$ 0.019  \\
                     & 5 & 15.805 $\pm$ 0.032 & 15.175 $\pm$ 0.015  \\
                     & 6 & 15.667 $\pm$ 0.013 & 14.649 $\pm$ 0.011  \\
                     & 7 & 14.811 $\pm$ 0.003 & 14.116 $\pm$ 0.010  \\
                     & 8 & 13.869 $\pm$ 0.013 & 13.275 $\pm$ 0.016  \\
                     & 9 & 14.626 $\pm$ 0.016 & 13.385 $\pm$ 0.017  \\
                     & 10& 16.416 $\pm$ 0.012 & 15.508 $\pm$ 0.014  \\
                     & 11& 15.656 $\pm$ 0.016 & 14.692 $\pm$ 0.015  \\
                     & 12& 15.956 $\pm$ 0.027 & 15.237 $\pm$ 0.022  \\
SBS 1116+583A        & 1 & 17.392 $\pm$ 0.004 & 17.161 $\pm$ 0.046  \\
                     & 2 & 14.242 $\pm$ 0.006 & 13.775 $\pm$ 0.014  \\
                     & 3$^{\dagger}$ & 17.167 & 16.311 \\
                     & 4$^{\dagger}$ & 17.166 & 16.537 \\
                     & 5 & 17.036 $\pm$ 0.008 & 16.468 $\pm$ 0.005  \\
Arp 151              & 1 & 14.300 $\pm$ 0.079 & 13.166 $\pm$ 0.078  \\
                     & 2 & 15.487 $\pm$ 0.084 & 14.814 $\pm$ 0.082  \\
                     & 3 & 15.496 $\pm$ 0.072 & 15.442 $\pm$ 0.083  \\
                     & 4 & 15.674 $\pm$ 0.073 & 15.027 $\pm$ 0.081  \\
                     & 5 & 15.610 $\pm$ 0.069 & 15.107 $\pm$ 0.084  \\
                     & 6 & 15.054 $\pm$ 0.074 & 14.306 $\pm$ 0.086  \\
                     & 7 & 15.417 $\pm$ 0.076 & 14.720 $\pm$ 0.087  \\
Mrk 1310             & 1 & 16.780 $\pm$ 0.003 & 15.752 $\pm$ 0.002  \\
Mrk 202              & 1$^{\dagger}$ & 17.221 & 16.429 \\
                     & 2$^{\dagger}$ & 17.907 & 17.412 \\
                     & 3 & 17.253 $\pm$ 0.013 & 15.710 $\pm$ 0.019  \\
                     & 4 & 15.309 $\pm$ 0.008 & 14.560 $\pm$ 0.001  \\
NGC 4253             & 1 & 17.247 $\pm$ 0.115 & 15.848 $\pm$ 0.025  \\
                     & 2 & 16.157 $\pm$ 0.045 & 15.166 $\pm$ 0.014  \\
                     & 3 & 16.293 $\pm$ 0.037 & 15.602 $\pm$ 0.021  \\
                     & 4 & 17.236 $\pm$ 0.082 & 16.323 $\pm$ 0.029  \\
                     & 5 & 14.708 $\pm$ 0.023 & 14.218 $\pm$ 0.021  \\
                     & 6 & 16.483 $\pm$ 0.013 & 15.599 $\pm$ 0.035  \\
                     & 7 & 16.833 $\pm$ 0.042 & 15.943 $\pm$ 0.031  \\
                     & 8 & 16.153 $\pm$ 0.045 & 16.195 $\pm$ 0.019  \\
NGC 4748             & 1 & 16.340 $\pm$ 0.020 & 15.660 $\pm$ 0.010  \\
                     & 2 & 17.130 $\pm$ 0.010 & 16.650 $\pm$ 0.010  \\
                     & 3$^{\dagger}$ & 17.430 & 16.854 \\
IC 4218              & 1 & 16.760 $\pm$ 0.010 & 15.990 $\pm$ 0.010  \\
                     & 2 & 18.290 $\pm$ 0.020 & 17.680 $\pm$ 0.010  \\
                     & 3$^{\dagger}$ &  17.480 & 17.060 \\
                     & 4$^{\dagger}$ &  18.750 & 18.008 \\
MCG-06-30-15         & 1 & 16.128 $\pm$ 0.014 & 15.163 $\pm$ 0.010  \\
NGC 5548             & 1 & 14.422 $\pm$ 0.006 & 13.782 $\pm$ 0.003  \\
                     & 2 & 13.742 $\pm$ 0.004 & 13.192 $\pm$ 0.002  \\
Mrk 290              & 1 & 18.298 $\pm$ 0.015 & 17.648 $\pm$ 0.001  \\
                     & 2 & 18.435 $\pm$ 0.007 & 17.302 $\pm$ 0.003  \\
                     & 3 & 18.849 $\pm$ 0.001 & 17.395 $\pm$ 0.047  \\
                     & 4 & 17.471 $\pm$ 0.013 & 16.513 $\pm$ 0.016  \\
                     & 5 & 17.213 $\pm$ 0.021 & 16.636 $\pm$ 0.002  \\
                     & 6 & 16.107 $\pm$ 0.001 & 15.474 $\pm$ 0.003  \\
                     & 7$^{\dagger}$ &  15.667 & 15.480 \\
IC 1198              & 1 & 14.855 $\pm$ 0.009 & 14.146 $\pm$ 0.007  \\
                     & 2 & 15.773 $\pm$ 0.018 & 15.146 $\pm$ 0.010  \\
                     & 3 & 16.889 $\pm$ 0.016 & 16.385 $\pm$ 0.044  \\
                     & 4 & 15.437 $\pm$ 0.019 & 14.537 $\pm$ 0.013  \\
                     & 5 & 15.893 $\pm$ 0.010 & 14.860 $\pm$ 0.045  \\
                     & 6 & 15.705 $\pm$ 0.016 & 14.856 $\pm$ 0.016  \\
                     & 7 & 16.263 $\pm$ 0.035 & 15.547 $\pm$ 0.011  \\
NGC 6814             & 1 & 16.779 $\pm$ 0.010 & 15.690 $\pm$ 0.005  \\

\enddata

\tablecomments{The quoted error is the standard deviation of all the
measurements for the comparison star taken during the photometric
night, and represents an estimate of the accuracy of the photometric
calibration. The comparison stars marked with a dagger were not
observed on a photometric night, and therefore were not directly
calibrated using \citet{Landolt_1992} standard stars. Instead, these 
stars were calibrated based on the other comparison stars for the AGN.}

\end{deluxetable}

\subsection{MAGNUM Photometry}
\label{subsec:magnumphot}

The photometry for the MAGNUM objects was carried out in a manner
similar to the photometry for the KAIT, Tenagra, and P60 objects. For
each of the MAGNUM images, the IRAF task {\tt phot} within the {\tt
daophot} package was used to measure the flux of the AGN and the sky
background through a circular aperture and a surrounding annulus. For
Mrk 1310, MCG-06-30-15, and NGC 6814, the AGN flux was compared to a
single star that fell within the MAGNUM field of view. For NGC 5548,
the AGN flux was compared to two stars which did not fall within the
same field of view, but were observed in an alternating pattern with
the AGN as described by \cite{Suganuma_2006}. The AGN magnitude was
found by averaging over the multiple exposures taken throughout the
night.

The size of the aperture and sky annulus for the MAGNUM objects was
set based upon past photometry of NGC 5548, and we did not experiment
with a range of aperture sizes as was discussed above for the KAIT,
Tenagra, and P60 objects. Following \cite{Suganuma_2006}, we used an
aperture size of 4\farcs15 and a sky annulus of 5\farcs54--6\farcs92
for NGC 5548. For the remaining MAGNUM objects, we reduced the
aperture radius to 2\farcs08 because the comparison stars were located
in the same field of view as the object. The sky annulus for Mrk 1310
and NGC 6814 was the same size as the one used for NGC 5548. Since the
nucleus of MCG-06-30-15 and the comparison star were located close
together, about 5\arcsec\ apart, we enlarged the sky annulus for this
object to 9\farcs13--10\farcs52.

As a final step, the $B$ and $V$-band light curves were flux
calibrated using \citet{Landolt_1992} standard stars as discussed in
\S \ref{sec:obs}. Unlike the flux calibration for the KAIT, Tenagra,
and P60 images, we did not fit the color term, but the correction is
expected to be small. We found the difference between the instrumental
and calibrated magnitude for each of the comparison stars, and applied
the average offset to the AGN light curve. The accuracy of the
photometric zeropoint calibration for the MAGNUM objects is estimated
to be better than $\sim 0.14$ mag.

\subsection{Host-Galaxy Subtraction}
\label{subsec:hostgalsub}

The $B$- and $V$-band photometric light curves include a combination
of AGN light and flux from the host-galaxy starlight, which dilutes
the observed AGN variations in the light curve. We removed host-galaxy
starlight by selecting the best image in each band and using the
two-dimensional image decomposition package Galfit \citep{Peng_2002}.

\begin{figure*}
\begin{center}
\epsscale{0.23}
\plotone{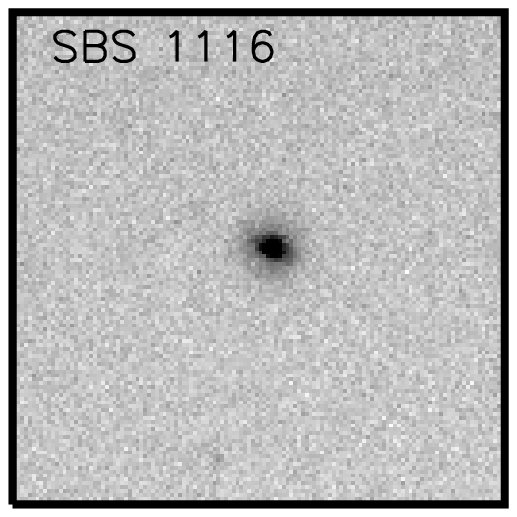}
\plotone{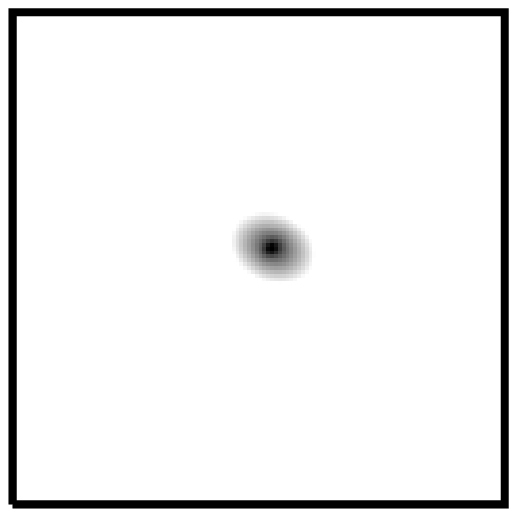}
\plotone{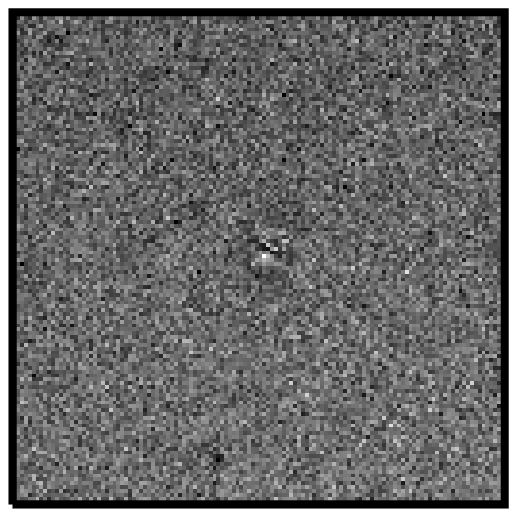} \\
\vspace{0.3cm}
\plotone{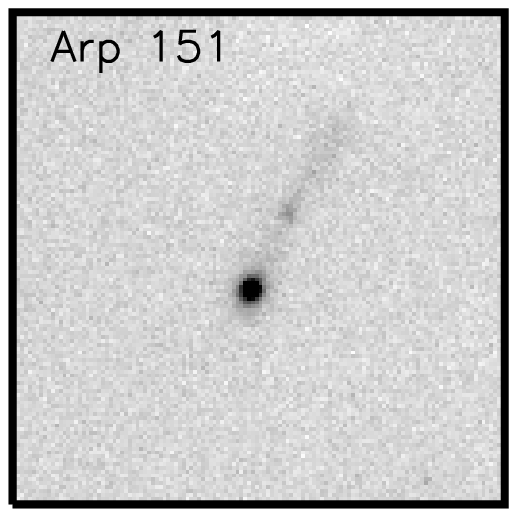}
\plotone{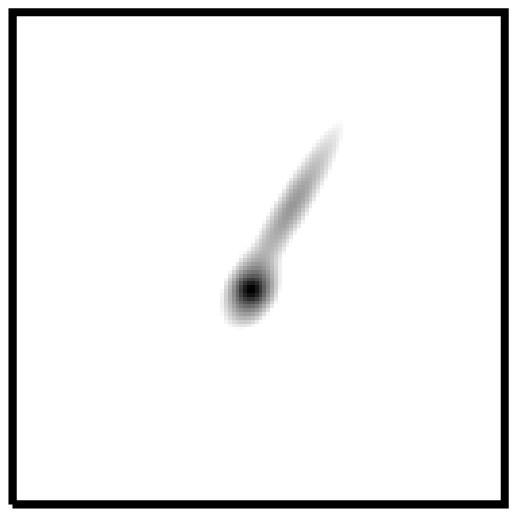}
\plotone{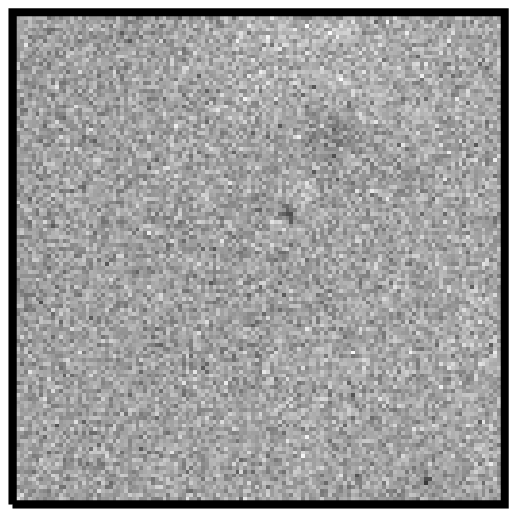} \\
\vspace{0.3cm}
\plotone{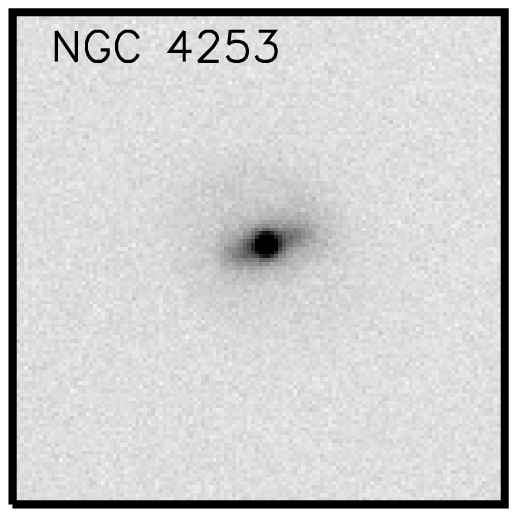}
\plotone{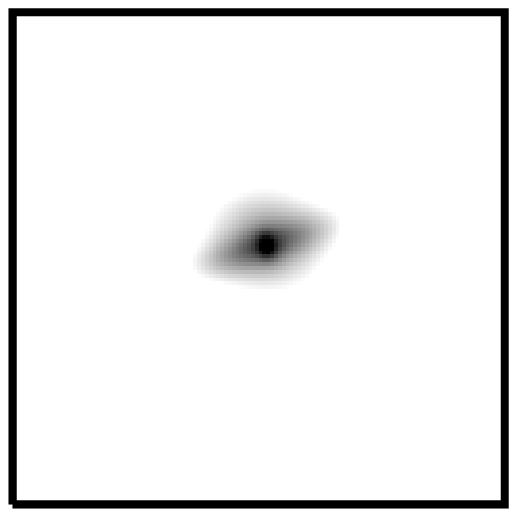}
\plotone{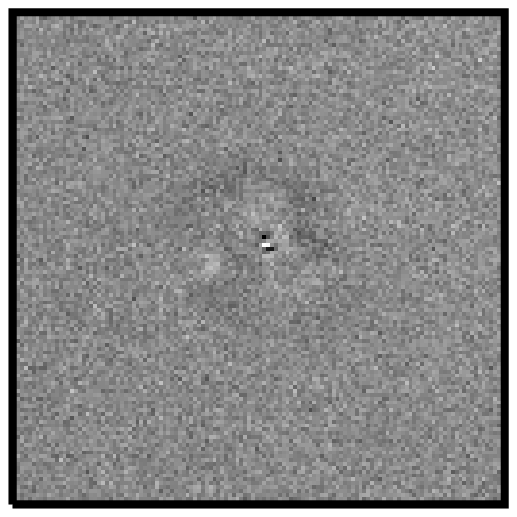} \\
\vspace{0.3cm}
\plotone{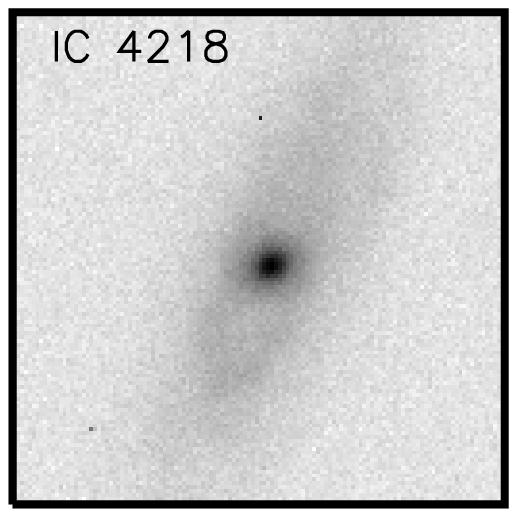}
\plotone{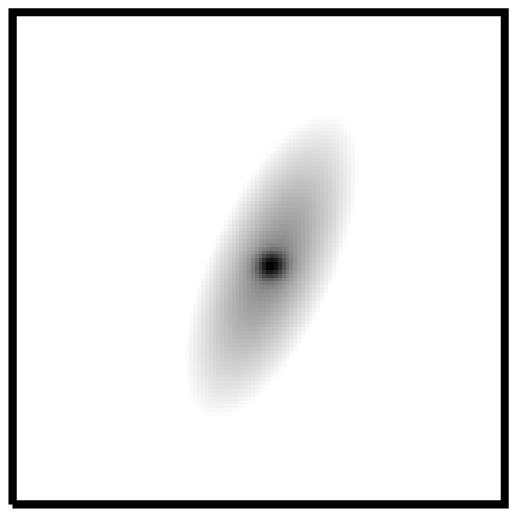}
\plotone{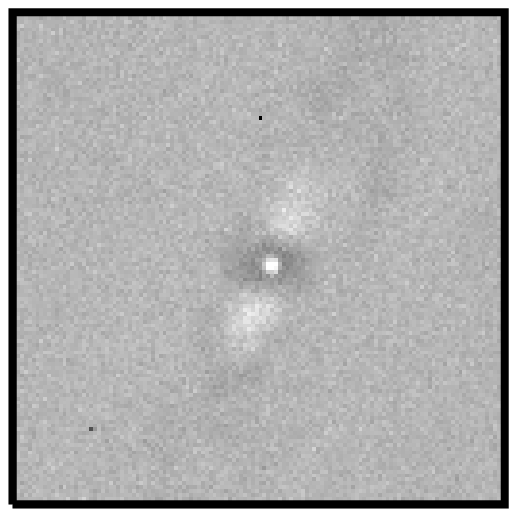}
\caption{From left to right: $V$-band image of AGN and host galaxy;
the best-fit model from Galfit consisting of a PSF for the nucleus and
an exponential function for the disk (and an additional exponential
function for the bar in NGC 4253 and for the tidal tail in Arp 151);
residuals of the fit. \label{fig:galfitex}}
\end{center}
\end{figure*}

We first fit a Moffat function to a nearby, bright comparison star. We
used the resultant model as the input point-spread function (PSF). We
focused on a subsection of the image centered on the galaxy and used
Galfit to fit a single exponential function for the disk, a PSF for
the nucleus, and a sky component. Due to the seeing effects and the
low resolution of the ground-based images, we are unable to properly
disentangle the bulge of the galaxy from the nucleus (or even discern
whether a bulge is present). Consequently, we do not include a bulge
component in the model. More complete modeling will be done in the
future based on Cycle 17 \emph{Hubble Space Telescope} (\emph{HST})
high-resolution images (GO-11662, PI: Bentz). Two exponential
functions were needed in the cases of NGC 4253 and IC 1198 to account
for both the disk and the bar in these objects. Also, Arp 151 and NGC
4748 were fit with two exponential functions. The second exponential
function was used to fit the tidal tail seen in Arp 151 and the disk
of the second galaxy near NGC 4748. Stars projected close to the
galaxy were masked out during the fitting process. In Figure
\ref{fig:galfitex}, we present a few examples of the Galfit models.


We then measured the galaxy magnitude from the model through the
appropriate aperture. We converted the data points in the light curve
from magnitudes to flux units, subtracted the constant galaxy flux
from each of the points in the light curve, and converted back to
magnitudes. In Table \ref{tab:hostgal}, we list the host-galaxy
magnitude measured through the aperture in each band.


\begin{deluxetable}{lccc}
\tabletypesize{\scriptsize} 
\tablewidth{0pt} 
\tablecaption{Host-Galaxy Magnitude \label{tab:hostgal}} 
\tablehead{
\colhead{Object} & 
\colhead{Aperture Radius} & 
\colhead{$B$} &
\colhead{$V$} \\ 
\colhead{} &
\colhead{(\arcsec)} & 
\colhead{(mag)} &
\colhead{(mag)} \\
\colhead{(1)} &
\colhead{(2)} & 
\colhead{(3)} &
\colhead{(4)}
}

\startdata

Mrk 142             &  4.35  & 18.47  & 17.28  \\
SBS 1116+583A       &  3.20  & 16.92  & 16.07  \\
Arp 151             &  4.35  & 17.30  & 16.30  \\
Mrk 1310            &  2.08  & 17.56  & 16.71  \\
Mrk 202             &  3.20  & 17.83  & 17.22  \\
NGC 4253            &  4.35  & 16.08  & 15.36  \\
NGC 4748            &  2.65  & 17.50  & 16.66  \\
IC 4218             &  2.65  & 17.90  & 16.96  \\
MCG-06-30-15        &  2.08  & 16.72  & 15.82  \\
NGC 5548            &  4.15  & 16.17  & 14.96  \\
Mrk 290             &  3.20  & 18.36  & 17.02  \\
IC 1198             &  4.35  & 16.89  & 15.94  \\
NGC 6814            &  2.08  & 16.64  & 15.61  \\

\enddata

\tablecomments{Columns 3 and 4 provide the estimated $B$- and $V$-band
host-galaxy magnitude measured through the aperture given in column
2. Host-galaxy contributions were estimated based on the simple Galfit
models described in \S \ref{subsec:hostgalsub}.}

\end{deluxetable}

The simple host-galaxy subtraction method was applied to all of the
light curves except those for NGC 5548. \cite{Suganuma_2006} provide a
measurement of the $B$- and $V$-band host-galaxy flux of NGC 5548
though an aperture with a radius of 4\farcs15. They additionally
supply a correction to the host-galaxy flux which depends on the
seeing. Since we used the same observational setup for our work, we
used the results from \cite{Suganuma_2006} to remove the host-galaxy
starlight of NGC 5548. We determined the median seeing over all the
images in each band and applied the seeing correction to the $B$- and
$V$-band host-galaxy flux. The host-galaxy offset was then subtracted
from each of the points on the light curve.

\subsection{Error Estimation}
\label{subsec:errors}

The uncertainty in the measurements was first calculated by the IRAF
package {\tt daophot}, which uses photon statistics and accounts for
the total number of pixels in the aperture and sky annulus, as well as
the read noise and gain of the detector. However, using photon
statistics alone can underestimate the error. Both seeing variations,
which affect the extended host galaxy differently than a point source,
and large color differences between the AGN and the comparison stars,
can increase the measurement errors. Therefore, we applied a secondary
uncertainty estimation method. We selected a relatively flat or slowly
varying portion of each light curve and calculated the average
difference between pairs of points closely spaced in time. This method
assumes that the objects do not exhibit microvariability [but
\cite{Klimek_2004} found that narrow-line Seyfert 1 galaxies can show
signs of variability over timescales of hours]. For each point on the
light curve, we adopted the larger of the two uncertainty
measurements. Most often, the secondary method produced the larger
error measurement. However, in instances where there was poor seeing
or bad weather, the photon-counting error was greater. The uncertainty
in the photometric measurements affects our ability to determine the
average time lag between continuum and emission-line
variation. Choosing the larger error will result in a more
conservative estimate of the time lag and its associated
uncertainties.

\section{Results}
\label{sec:results}

\begin{figure}
\begin{center}
\plotone{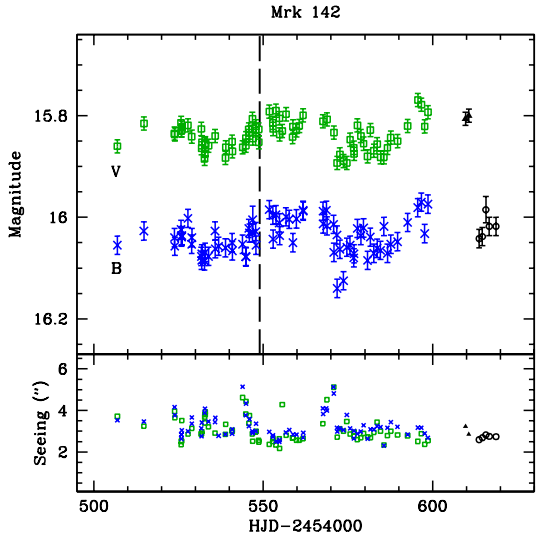}
\caption{$B$-band and $V$-band light curves of the Mrk 142 nucleus
for the period between 2008 February and 2008 May, along with the
measured seeing plotted as a function of time. The light curves have
been corrected for Galactic extinction using the values given in Table
\ref{tab:galprop}. Blue crosses represent $B$-band measurements and
green open squares represent the $V$-band measurements. On several
nights, Mrk 142 was observed by KAIT in place of the primary
telescope. The KAIT observations in the $B$ and $V$ bands are shown in
the black open circles and the black filled triangles,
respectively. The dashed vertical line indicates when the
spectroscopic monitoring began. \label{fig:mrk142ltcurves}}
\end{center}
\end{figure}

\begin{figure}
\begin{center}
\plotone{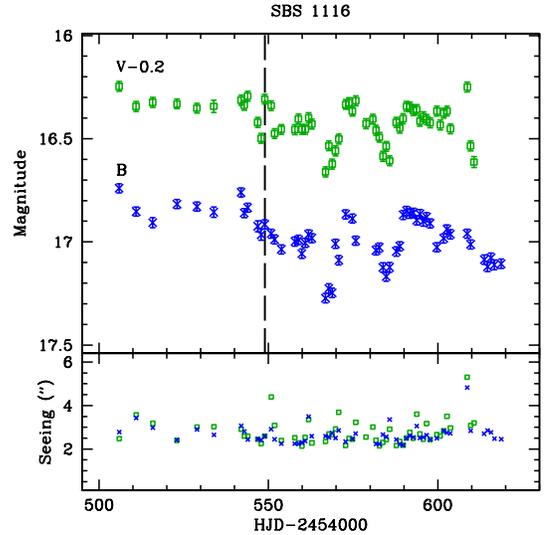}
\caption{SBS 1116; see Figure \ref{fig:mrk142ltcurves} for
description. \label{fig:sbs1116ltcurves}}
\end{center}
\end{figure}

\begin{figure}
\begin{center}
\plotone{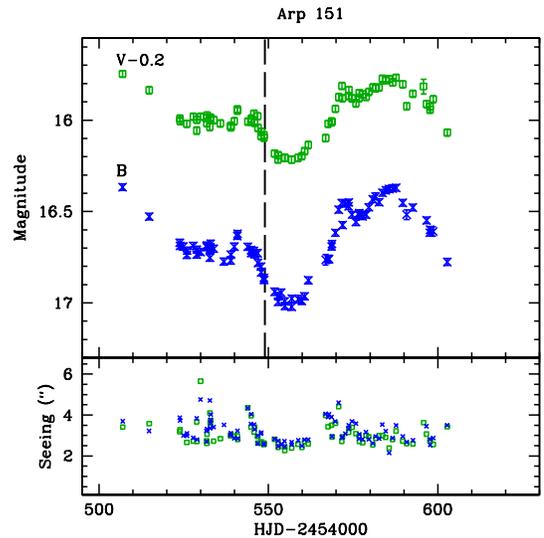}
\caption{Arp 151; see Figure \ref{fig:mrk142ltcurves} for
description. \label{fig:arp151ltcurves}}
\end{center}
\end{figure}

\begin{figure}
\begin{center}
\plotone{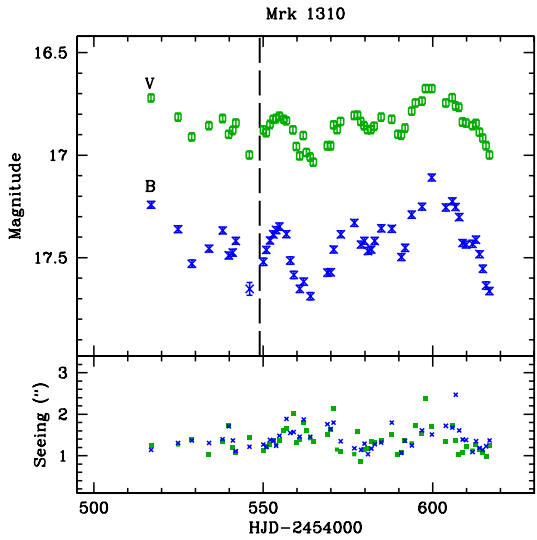}
\caption{Mrk 1310; see Figure \ref{fig:mrk142ltcurves} for
description. \label{fig:mrk13101ltcurves}}
\end{center}
\end{figure}

\begin{figure}
\begin{center}
\plotone{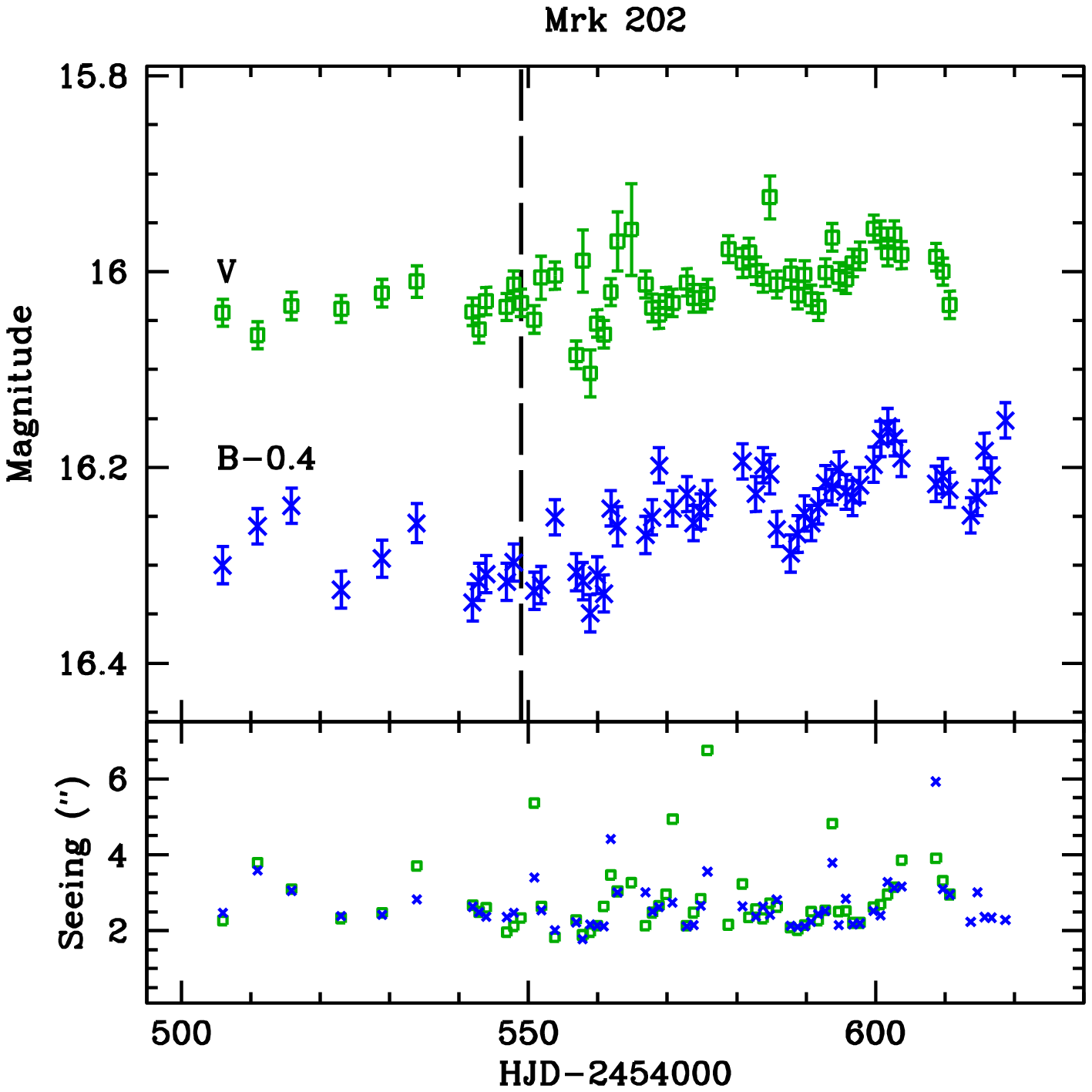}
\caption{Mrk 202; see Figure \ref{fig:mrk142ltcurves} for
description. \label{fig:mrk202ltcurves}}
\end{center}
\end{figure}

\begin{figure}
\begin{center}
\plotone{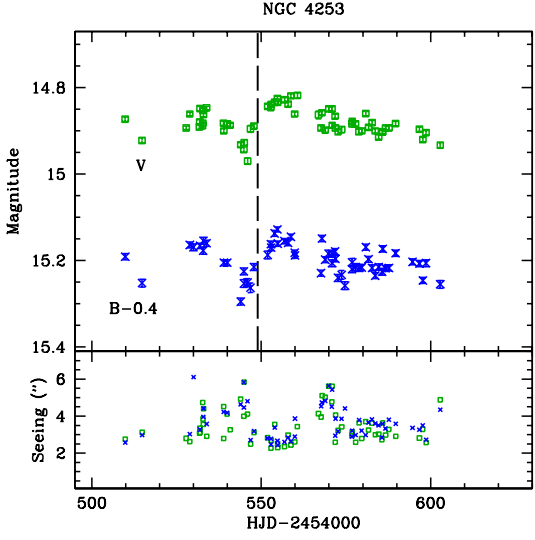}
\caption{NGC 4253; see Figure \ref{fig:mrk142ltcurves} for
description. \label{fig:mrk766ltcurves}}
\end{center}
\end{figure}

\begin{figure}
\begin{center}
\plotone{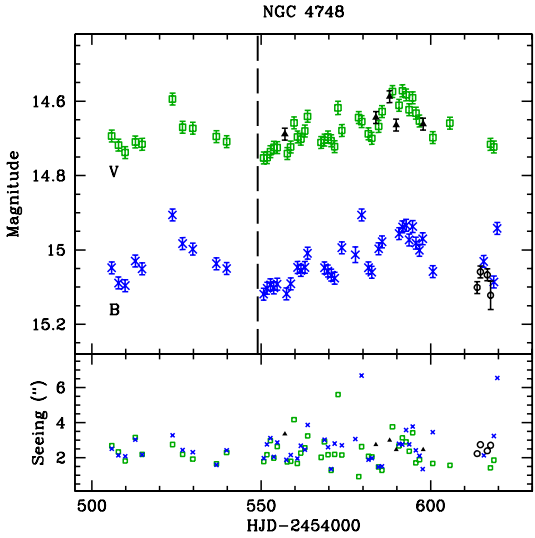}
\caption{NGC 4748; see Figure \ref{fig:mrk142ltcurves} for
description. \label{fig:ngc4748ltcurves}}
\end{center}
\end{figure}

\begin{figure}
\begin{center}
\plotone{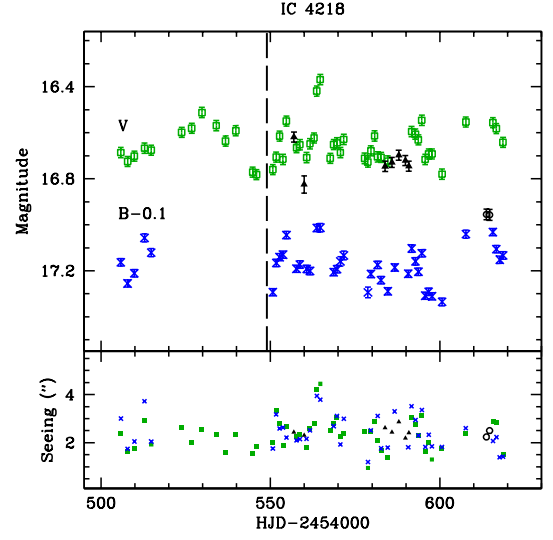}
\caption{IC 4218; see Figure \ref{fig:mrk142ltcurves} for
description. \label{fig:ic4218ltcurves}}
\end{center}
\end{figure}

\begin{figure}
\begin{center}
\plotone{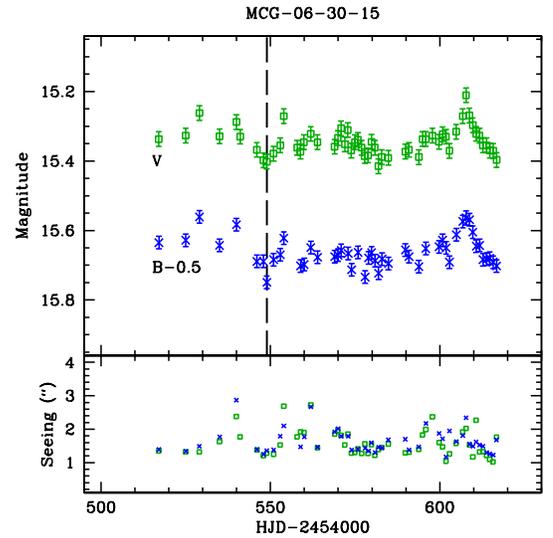}
\caption{MCG-06-30-15; see Figure \ref{fig:mrk142ltcurves} for
description. \label{fig:mcg06ltcurves}}
\end{center}
\end{figure}

\begin{figure}
\begin{center}
\plotone{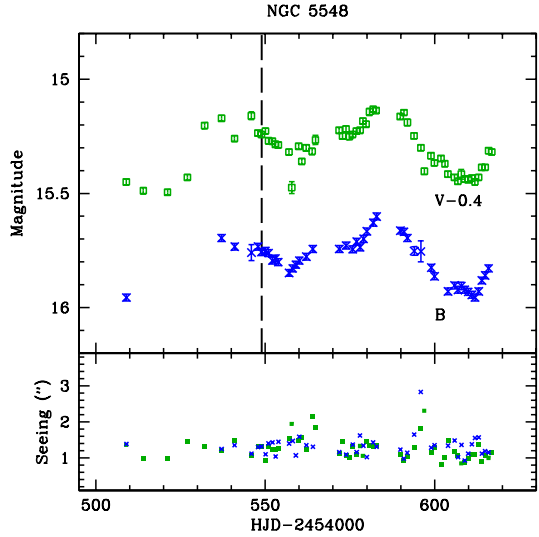}
\caption{NGC 5548; see Figure \ref{fig:mrk142ltcurves} for
description. \label{fig:ngc5548ltcurves}}
\end{center}
\end{figure}

\begin{figure}
\begin{center}
\plotone{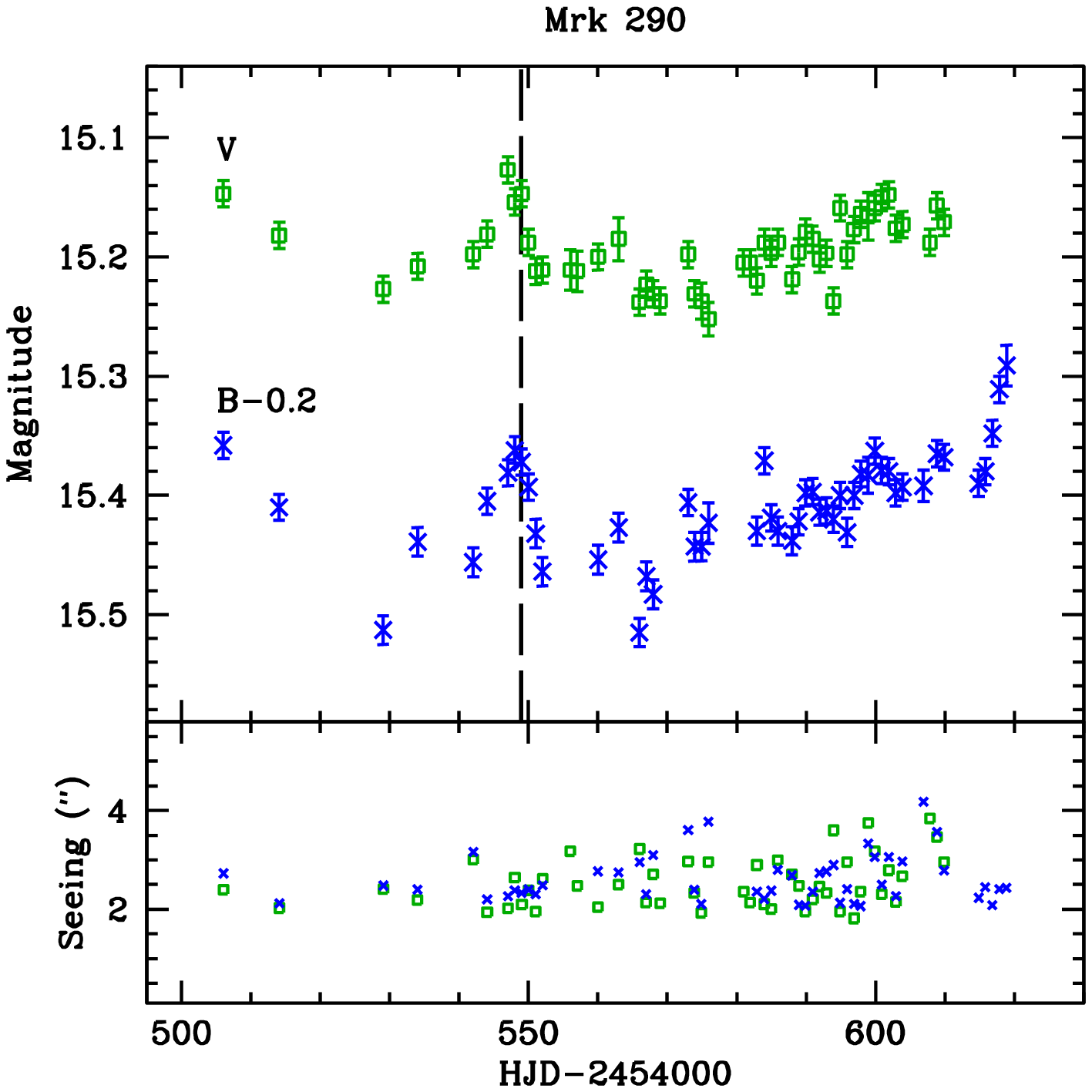}
\caption{Mrk 290; see Figure \ref{fig:mrk142ltcurves} for
description. \label{fig:mrk290ltcurves}}
\end{center}
\end{figure}

\begin{figure}
\begin{center}
\plotone{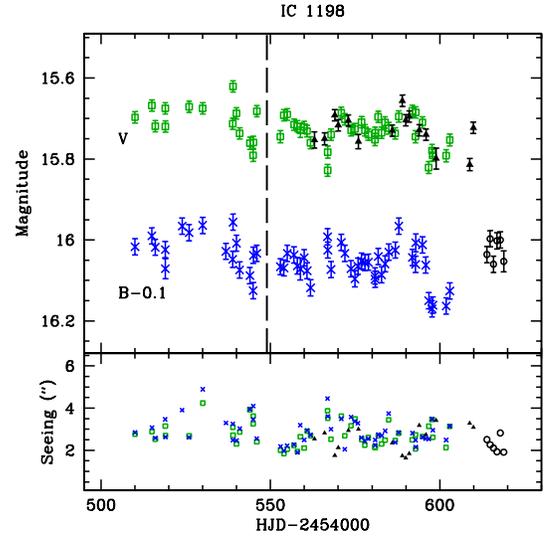}
\caption{IC 1198; see Figure \ref{fig:mrk142ltcurves} for
description. \label{fig:mrk871ltcurves}}
\end{center}
\end{figure}

\begin{figure}
\begin{center}
\plotone{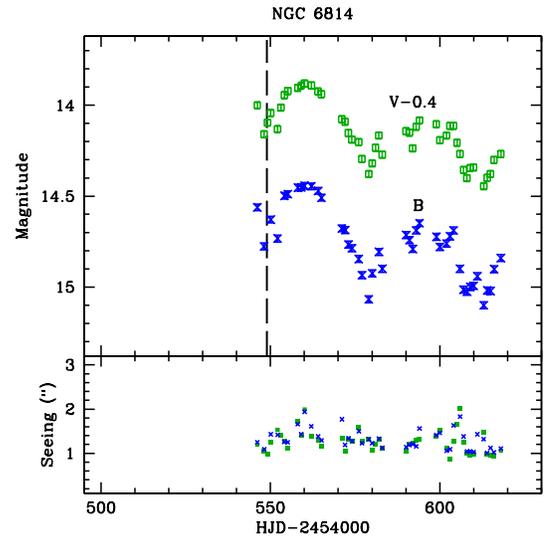}
\caption{NGC 6814; see Figure \ref{fig:mrk142ltcurves} for
description. \label{fig:ngc6814ltcurves}}
\end{center}
\end{figure}

We present the nuclear $B$- and $V$-band light curves for the full
sample, along with the estimated seeing for each image, in Figures
\ref{fig:mrk142ltcurves}--\ref{fig:ngc6814ltcurves} and in Table
\ref{tab:ltcurvesall_stub}. These light curves include the host-galaxy
pedestal subtraction and have been corrected for Galactic extinction
using the values given in Table \ref{tab:galprop}.


\begin{deluxetable*}{lcccccccccc}
\tabletypesize{\scriptsize} 
\tablewidth{0pt} 
\tablecaption{Photometry \label{tab:ltcurvesall_stub}} 
\tablehead{
\colhead{} &
\colhead{} &
\multicolumn{4}{c}{$B$ Band} &&
\multicolumn{4}{c}{$V$ Band} \\
\cline{3-6}\cline{8-11} \\
\colhead{Object} &
\colhead{UT} & 
\colhead{HJD} & 
\colhead{$B$} & 
\colhead{error} & 
\colhead{FWHM} &
\colhead{} & 
\colhead{HJD} &
\colhead{$V$} &
\colhead{error} &
\colhead{FWHM} \\ 
\colhead{} &
\colhead{Date} &
\colhead{$-$2454000} & 
\colhead{(mag)} & 
\colhead{(mag)} &
\colhead{(arcsec)} &
\colhead{} &
\colhead{$-$2454000} & 
\colhead{(mag)} & 
\colhead{(mag)} &
\colhead{(arcsec)}
}

\startdata

Mrk 142  &  2008 Feb 10  &  506.9393  &  16.055  &  0.018  &  3.5  &&  506.9473  &  15.860  &  0.013  &  3.7  \\
Mrk 142  &  2008 Feb 18  &  514.7661  &  16.027  &  0.018  &  3.5  &&  514.7742  &  15.815  &  0.013  &  3.2  \\
Mrk 142  &  2008 Feb 27  &  523.8112  &  16.040  &  0.018  &  4.2  &&  523.8193  &  15.835  &  0.013  &  4.0  \\
Mrk 142  &  2008 Feb 27  &  523.9006  &  16.057  &  0.018  &  3.8  &&  523.9085  &  15.837  &  0.013  &  3.6  \\ 
Mrk 142  &  2008 Feb 29  &  525.7402  &  16.024  &  0.018  &  2.8  &&  525.7496  &  15.814  &  0.013  &  2.4  \\

\enddata

\tablecomments{Table \ref{tab:ltcurvesall_stub} is published in its
entirety in the electronic edition of the \emph{Astrophysical Journal
Supplement}. A portion of Table \ref{tab:ltcurvesall_stub} is shown
here for a guidance regarding the form and content.}

\end{deluxetable*}


\begin{deluxetable*}{lccccccccc}
\tabletypesize{\scriptsize} 
\tablewidth{0pt} 
\tablecaption{Variability Statistics \label{tab:variablitystats}} 
\tablehead{
\colhead{} &
\multicolumn{4}{c}{$B$ Band} &&
\multicolumn{4}{c}{$V$ Band} \\
\cline{2-5}\cline{7-10} \\
\colhead{Object} & 
\colhead{$<f>$} & 
\colhead{$\sigma$} & 
\colhead{$F_\mathrm{var}$} & 
\colhead{$R_\mathrm{max}$} &
\colhead{} &
\colhead{$<f>$} & 
\colhead{$\sigma$} &
\colhead{$F_\mathrm{var}$} &
\colhead{$R_\mathrm{max}$} \\ 
\colhead{} &
\colhead{(mJy)} & 
\colhead{(mJy)} & 
\colhead{} &
\colhead{} &
\colhead{} &
\colhead{(mJy)} & 
\colhead{(mJy)} & 
\colhead{} &
\colhead{} \\
\colhead{(1)} &
\colhead{(2)} & 
\colhead{(3)} & 
\colhead{(4)} &
\colhead{(5)} &
\colhead{} &
\colhead{(6)} & 
\colhead{(7)} & 
\colhead{(8)} &
\colhead{(9)}
}

\startdata

Mrk 142       &  1.57  &  0.05  &  0.03  &  1.17  &&  1.64  &  0.04  &  0.02  &  1.12  \\
SBS 1116+583A &  0.66  &  0.07  &  0.10  &  1.63  &&  0.80  &  0.07  &  0.08  &  1.47  \\
Arp 151       &  0.89  &  0.14  &  0.16  &  1.83  &&  1.22  &  0.14  &  0.11  &  1.54  \\
Mrk 1310      &  0.44  &  0.05  &  0.12  &  1.70  &&  0.65  &  0.05  &  0.07  &  1.39  \\
Mrk 202       &  0.89  &  0.04  &  0.04  &  1.20  &&  1.40  &  0.04  &  0.03  &  1.18  \\
NGC 4253      &  2.35  &  0.08  &  0.03  &  1.17  &&  3.97  &  0.11  &  0.03  &  1.15  \\
NGC 4748      &  3.96  &  0.22  &  0.05  &  1.22  &&  4.80  &  0.22  &  0.04  &  1.18  \\
IC 4218       &  0.51  &  0.05  &  0.09  &  1.42  &&  0.78  &  0.06  &  0.08  &  1.52  \\
MCG-06-30-15  &  1.40  &  0.06  &  0.04  &  1.19  &&  2.59  &  0.10  &  0.03  &  1.21  \\ 
NGC 5548      &  1.97  &  0.17  &  0.08  &  1.39  &&  1.87  &  0.18  &  0.09  &  1.40  \\
Mrk 290       &  2.33  &  0.09  &  0.04  &  1.23  &&  2.98  &  0.08  &  0.02  &  1.12  \\    
IC 1198       &  1.42  &  0.06  &  0.04  &  1.22  &&  1.82  &  0.07  &  0.03  &  1.21  \\
NGC 6814      &  5.15  &  0.92  &  0.18  &  1.83  &&  5.40  &  0.79  &  0.14  &  1.68  \\

\enddata

\tablecomments{Columns 2--5 list the mean flux ($<f>$), the rms
variation ($\sigma$), the normalized excess variance
($F_\mathrm{var}$), and the ratio of maximum to minimum fluxes
($R_\mathrm{max}$) for the $B$ band. Similarly, columns 6--9 list the
variability characteristics for the $V$ band. Magnitudes given in
Tables \ref{tab:ltcurvesall_stub}--18 were first converted to fluxes
using the calibration of Vega in the Johnson system before calculating
the variability statistics.}

\end{deluxetable*}

\subsection{Variability Characteristics}
\label{subsec:varchar}

In order to quantify the variability of the $B$- and $V$-band light
curves, in Table \ref{tab:variablitystats} we present variability
characteristics for each of the objects. We determine the mean flux
($\left<f\right>$), the root mean square (rms) deviation in the flux
($\sigma^2$), the ratio of the maximum to minimum flux
($R_\mathrm{max}$), and the normalized excess variance
($F_\mathrm{var}$). The normalized excess variance gives an estimate
of the intrinsic variability of the source corrected for measurement
uncertainties and normalized by the mean flux. It is defined according
to \cite{RodriguesPascual_1997} as

\begin{equation}
F_{\rm var}=\frac{\sqrt{\sigma^2-\delta^2}}{\left<f\right>} \;,
\label{mf_eq}
\end{equation}

{\noindent where $\delta^2$ is the average uncertainty of the
individual flux measurements.}

As expected, there is a wide range in the level of variability between
the 13 objects in the sample. About a third of the sample (including
Arp 151, Mrk 1310, NGC 5548, and NGC 6814) shows large variations of
about $\sim 0.5$ mag in both the $B$ and $V$ bands throughout the
course of the campaign.  However, most of the objects (including SBS
1116, Mrk 202, NGC 4253, NGC 4748 and Mrk 290) exhibit a single event
or an occasional moderate change in the continuum, where the magnitude
changes by $\sim 0.3$ mag.  Single coherent variations like those seen
at the beginning of the imaging campaign in NGC 4253 and Mrk 290
further emphasize the need for continuous daily monitoring lasting
several months, even for objects expected to have short time lags. The
remaining objects (Mrk 142, IC 4218, MCG-06-30-15, and IC 1198)
display only small variations (as little as $\sim 0.1$ mag) in the
continuum flux. In most cases, $F_\mathrm{var}$ and $R_\mathrm{max}$
are higher in the $B$ band than in the $V$ band, probably due to the
larger host-galaxy contribution that remains in the $V$ band.

\subsubsection{Contribution to the Variability from Broad Emission Lines}
\label{subsubsec:emissionlines}

For the objects in our sample, the H$\beta$ and H$\gamma$ emission
lines contribute of order 10\% to the measured AGN $B$-band flux. The
flux of both the broad H$\beta$ and H$\gamma$ lines varies throughout
the campaign, and thus some of the variability seen in the photometric
$B$-band light curve is due to the variability of the emission lines,
and is not part of the continuum variability. For example, if the
broad emission-line flux varies by 10\%, we expect the effect of the
broad emission-line variability on the variability of the photometric
light curve to be small, of order 1\%.

We estimated the size of this effect in Arp 151 by subtracting the
H$\beta$ and H$\gamma$ spectroscopic light curves (determined through
the spectroscopic campaign; see Paper III for the H$\beta$ light
curve) from the photometric $B$-band light curve. Before performing
the subtraction, we created a modified photometric light curve by
averaging together all photometric AGN measurements taken on a single
night, and removed measurements that were taken on nights when
spectroscopic measurements were not made as well. We converted the
photometric AGN measurements into flux units. We also scaled the
H$\beta$ and H$\gamma$ spectroscopic light curves so that the
measurements would represent the integrated flux of the emission lines
as measured through a Johnson $B$-band filter. Finally, we subtracted
the scaled H$\beta$ and H$\gamma$ spectroscopic light curves from the
modified photometric light curve, producing a light curve with
variability that is independent of the H$\beta$ and H$\gamma$
broad-line variability. We measured the variability characteristics of
the resultant light curve, finding that $R_\mathrm{max} = 1.80$ and
$F_\mathrm{var} = 0.19$. These values can be compared to the
variability characteristics, $R_\mathrm{max} = 1.79$ and
$F_\mathrm{var} = 0.19$, measured from the modified photometric light
curve. This verifies that the effect of the emission-line variability
on the variability of the photometric $B$-band light curve is small,
less than 1\%. Since Arp 151 is one of the most highly variable
objects in the sample, we expect the effect of emission-line
variability on the variability of the photometric light curves to be
smaller in the other objects. The broad-band $V$ filter contains a
similar contribution of flux from the H$\beta$ line (about 6\% for Arp
151) as the $B$-band filter, so the emission-line variations will not
contribute significantly to the $V$-band photometric variability.

\subsection{Cross-Correlation between $B$- and $V$-Band Light Curves}
\label{subsec:crosscorrelation}

One of the possible mechanisms driving the optical continuum
variability on short timescales is the reprocessing of higher energy
continuum fluctuations, such as X-ray variations, within the accretion
disk. The reprocessing scenario predicts that variations should be
seen first at short wavelengths, since the short-wavelength emission
originates from a region close to the nucleus, followed by variations
at longer wavelengths, since the long-wavelength emission arises from
portions of the accretion disk at larger radii. The time delay between
the short- and long-wavelength variations corresponds to the light
travel time between the emitting portions of the accretion
disk. Another possibility is that fluctuations in the accretion disk
drive the continuum variability. In this scenario, long-wavelength
variations should precede short-wavelength variations as the disk
fluctuations propagate inward on a viscous or thermal timescale. A
number of studies have searched for time delays between X-ray and
optical variations (e.g., \citealt{Nandra_1998, Maoz_2000,
Edelson_2000, Suganuma_2006, Arevalo_2008, Breedt_2009}), optical and
ultraviolet variations (e.g., \citealt{Collier_1998, Desroches_2006}),
optical and near-infrared variations (e.g., \citealt{Minezaki_2004,
Suganuma_2006}), and even between various optical bands or UV bands
(e.g., \citealt{Wanders_1997, Sergeev_2005, Doroshenko_2005,
Cackett_2007}). The past work has found differing results. In some
cases the variations seen at shorter wavelengths precede the
fluctuations at longer wavelengths. However, the converse has also
been observed, as well as no strong wavelength dependence, and a clear
picture has yet to emerge.

\begin{figure*}
\begin{center}
\epsscale{1.0}
\plotone{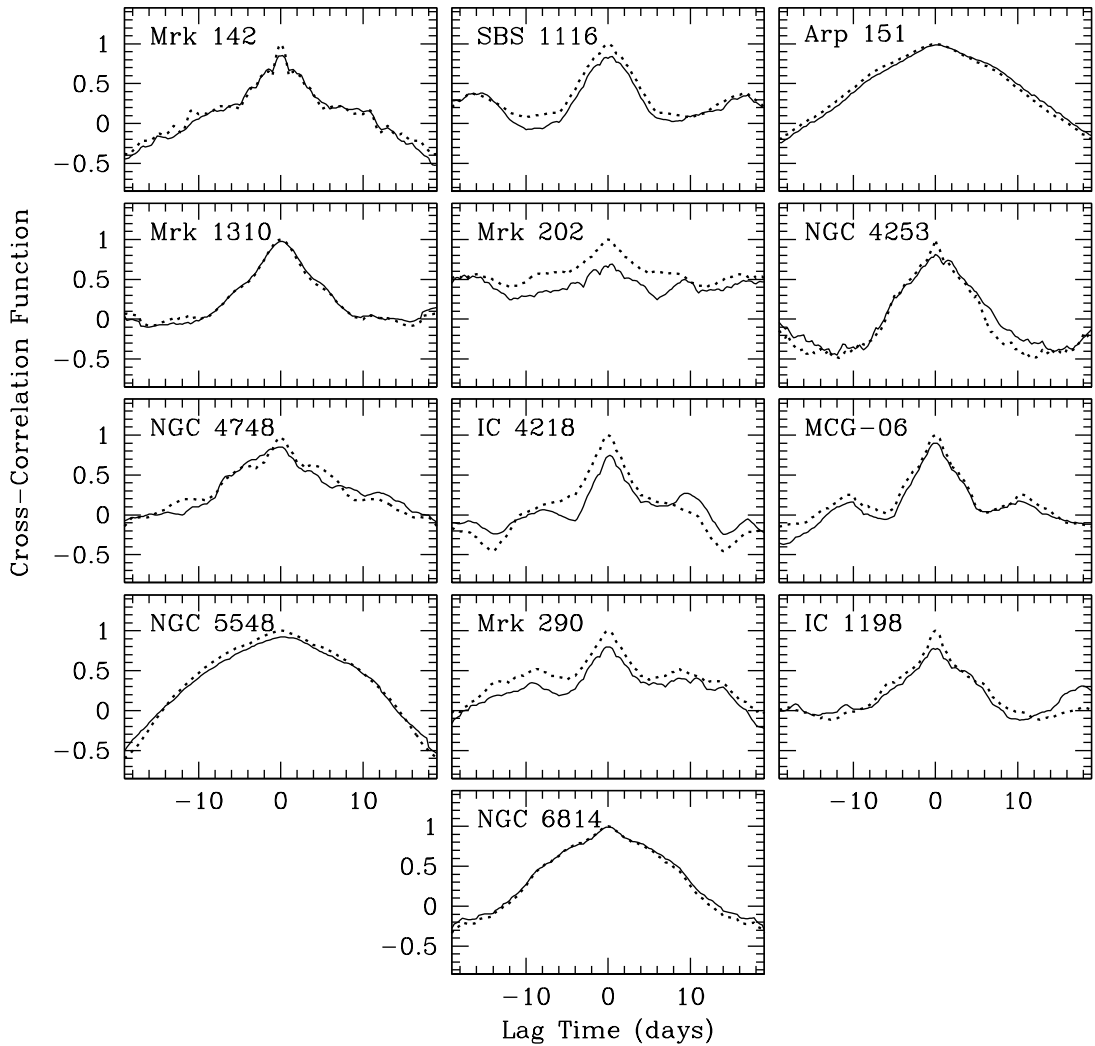}
\caption{Cross-correlation between the $B$- and $V$-band light curves
for each of the objects in the sample (shown in black). The $B$-band
autocorrelation function is shown by the dotted line for
comparison. \label{fig:ccf}}
\end{center}
\end{figure*}

For each of the objects in the sample, the variations in the $B$- and
$V$-band light curves are well correlated, and we can search for a
time delay between those variations. We measured the time delay
relative to the $B$-band light curve using the interpolation
cross-correlation method \citep{Gaskell_Sparke_1986,
Gaskell_Peterson_1987} with the \cite{White_Peterson_1994}
modifications, and found the lag at the cross-correlation function
(CCF) peak ($\tau_{\mathrm{peak}}$) and the CCF centroid
($\tau_{\mathrm{cent}}$). The CCF centroid is determined using points
near the CCF peak ($r_\mathrm{max}$) with the correlation coefficient
greater than or equal to 0.8$r_\mathrm{max}$. We list the values for
$\tau_{\mathrm{peak}}$ and $\tau_{\mathrm{cent}}$, along with
$r_\mathrm{max}$, for each of the objects in Table \ref{tab:ccf}. The
uncertainties were determined by using the Monte Carlo flux
randomization/random subset sampling (FR/RSS) technique of
\cite{Peterson_1998, Peterson_2004}. In Figure \ref{fig:ccf} we
present the CCFs for each of the objects, as well as the $B$-band
autocorrelation function.


\begin{deluxetable}{lccc}
\tabletypesize{\scriptsize} 
\tablewidth{0pt} 
\tablecaption{Cross-Correlation Results \label{tab:ccf}} 
\tablehead{
\colhead{Object} & 
\colhead{$r_\mathrm{max}$} & 
\colhead{$\tau_\mathrm{cent}$} & 
\colhead{$\tau_\mathrm{peak}$}  \\ 
\colhead{} & 
\colhead{} & 
\colhead{(days)} &
\colhead{(days)} \\
\colhead{(1)} & 
\colhead{(2)} & 
\colhead{(3)} &
\colhead{(4)}
}

\startdata

Mrk 142	     &  0.85  &  $-$0.01$^{+0.70}_{-0.62}$  &  0.00$^{+0.50}_{-0.50}$  \\
SBS1116+583A &  0.84  &   0.12$^{+1.52}_{-1.76}$  &  0.25$^{+0.25}_{-0.75}$  \\
Arp 151	     &  0.99  &   0.50$^{+0.62}_{-0.61}$  &  0.25$^{+0.25}_{-0.25}$  \\
Mrk 1310     &  0.98  &   0.13$^{+0.25}_{-0.25}$  &  0.25$^{+0.00}_{-0.25}$  \\
Mrk 202	     &  0.69  &   0.01$^{+1.36}_{-0.86}$  &  0.25$^{+0.50}_{-1.75}$  \\
NGC 4253     &  0.81  &   0.24$^{+0.49}_{-0.62}$  &  0.00$^{+0.25}_{-0.50}$  \\
NGC 4748     &  0.85  &  $-$0.85$^{+0.88}_{-0.83}$  & -0.25$^{+0.50}_{-0.50}$  \\
IC 4218	     &  0.75  &   0.26$^{+0.36}_{-0.37}$  &  0.25$^{+0.25}_{-0.25}$  \\
MCG-06-30-15 &  0.90  &   0.00$^{+0.48}_{-0.38}$  &  0.00$^{+0.50}_{-0.50}$  \\
NGC 5548     &  0.93  &  $-$0.02$^{+0.73}_{-0.81}$  &  0.25$^{+0.25}_{-0.75}$  \\
Mrk 290	     &  0.80  &  $-$0.00$^{+2.62}_{-3.35}$  &  0.00$^{+0.50}_{-0.50}$  \\
IC 1198	     &  0.78  &   0.12$^{+1.09}_{-0.71}$  &  0.25$^{+0.50}_{-0.75}$  \\	
NGC 6814     &  0.99  &   0.25$^{+0.63}_{-0.70}$  &  0.00$^{+0.25}_{-0.25}$

\enddata

\tablecomments{Columns 2--4 list the peak value of the CCF, the
position of the centroid of the CCF, and the position of the peak of
the CCF.}

\end{deluxetable}

We do not find significant positive or negative lags between the
variations of the $B$- and $V$-band light curves for any of the
objects in the sample. We measure both positive and negative lags that
range up to a few tenths of a day, but these measurements all have <
1$\sigma$ significance since they are well below our nightly sampling
cadence, and hence are all consistent with zero lag. These results are
not surprising since the $B$ and $V$ bands do not differ much in
wavelength, and previous photometric monitoring campaigns (e.g.,
\citealt{Sergeev_2005}) have found that lag times between $B$ and $V$
band variations are typically only a few tenths of a day. Due to the
near-simultaneity of variations between optical bands and between
optical and UV bands, past work has also disfavored variability models
in which photometric variability results from disk instabilities that
propagate inward at a speed much slower than the speed of light, and
instead has favored a reprocessing scenario (e.g.,
\citealt{Krolik_1991}, \citealt{Edelson_1996},
\citealt{Collier_1998}). Higher-frequency sampling is required to
carry out a more definitive search for time lags between variations in
different optical bands.


\subsection{Color Variability}
\label{subsec:colorvar}

\begin{figure*}
\begin{center}
\epsscale{1.0}
\plotone{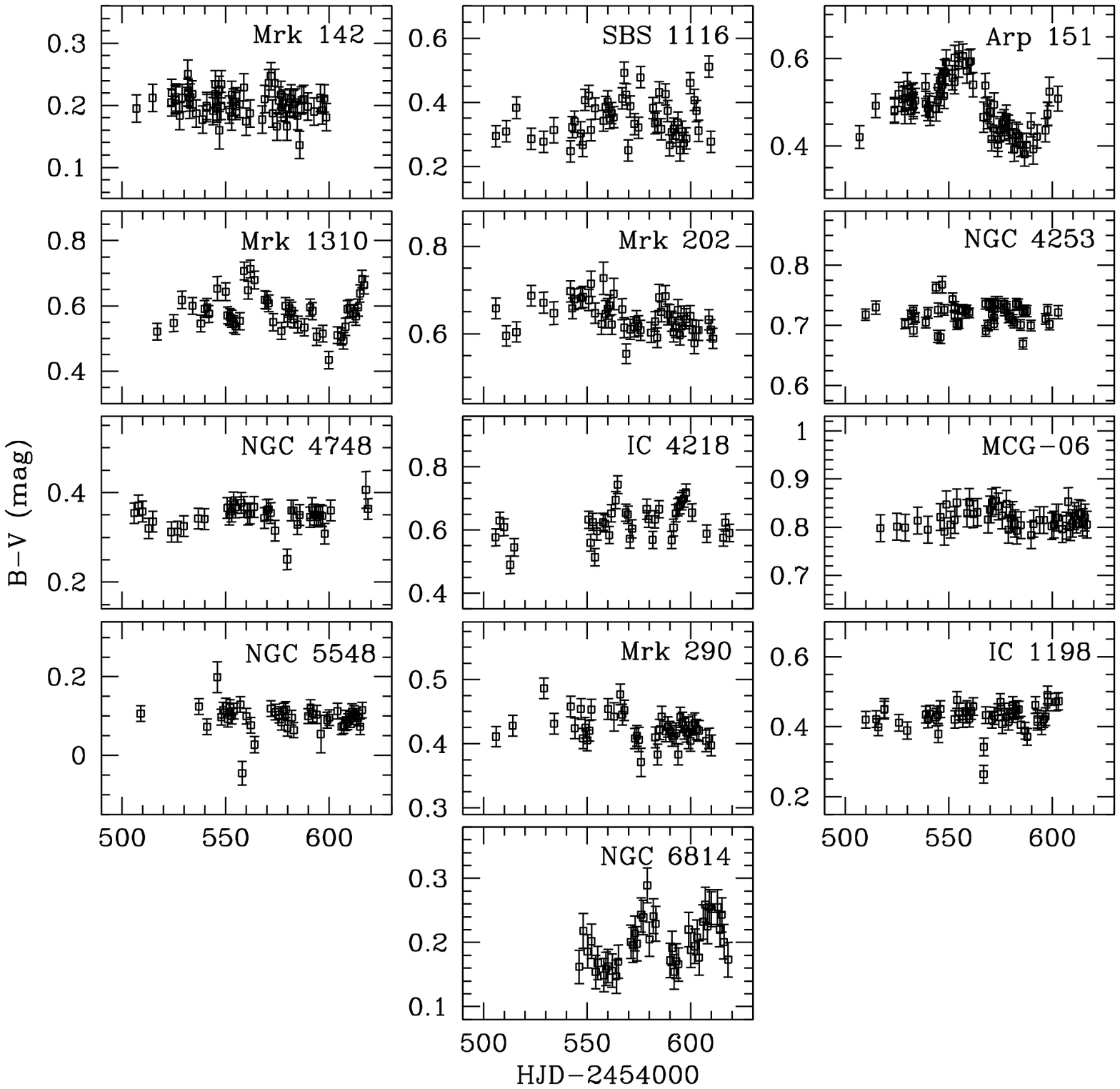}
\caption{$B-V$ color as a function of time for all of the objects in
the sample. The $B-V$ color was obtained after applying the simple
method of host-galaxy subtraction described in \S
\ref{subsec:hostgalsub}. \label{fig:colorvar}}
\end{center}
\end{figure*}

In addition to searching for time delays between variations in the $B$
and $V$ bands, we use the light curves to measure $B-V$ colors. Since
we employed a simple method of galaxy subtraction (\S
\ref{subsec:hostgalsub}) which did not account for a bulge component,
the $B-V$ color represents the color of the AGN and some portion of
the host galaxy. We present the results for all of the objects in
Figure \ref{fig:colorvar}.


We find that the color ranges from $B-V = 0.1$ mag to $B-V = 0.8$
mag. There is no convincing evidence for color variability, with the
exception of Arp 151, Mrk 1310, and NGC 6814. The color varied from a
maximum $B-V$ of 0.6 mag to a minimum $B-V$ of 0.4 mag for Arp 151,
from $B-V = 0.75$ mag to $B-V = 0.50$ mag for Mrk 1310, and from $B-V
= 0.28$ mag to $B-V = 0.15$ mag for NGC 6814. In each case, as the AGN
became fainter the $B-V$ color became redder. The simplest explanation
for this trend is that some of the host-galaxy starlight remains in
the $B$- and $V$-band light curves, which is the result of not
properly accounting for the bulge component. Therefore, as the AGN
becomes fainter, the host-galaxy contamination becomes more
significant, resulting in a redder color. The effect will be most
obvious for the highly variable objects, such as Arp 151, Mrk 1310,
and NGC 6814. The color variability seen in these objects shows the
limitations of our simple host-galaxy subtraction method.

Work by \cite{Sakata_2009} further explores color variability in a
sample of bright, highly variable Seyfert 1 galaxies using
well-sampled data from MAGNUM spanning seven years. They found that
the optical spectral shape is independent of flux variations. Also,
\cite{Woo_2007} studied a sample of moderate-luminosity AGNs and found
that the spectral shape remains the same on weekly timescales over a
rest-frame wavelength range of 2800--5200 \AA. The results found by
\cite{Sakata_2009} and \cite{Woo_2007} are consistent with the
constant colors found for our lower luminosity sample, and further
supports our interpretation that the color variability seen in Arp
151, Mrk 1310, and NGC 6814 is the result of a residual host-galaxy
contribution.

\section{Conclusions}
\label{sec:conclusions}

As part of the Lick AGN Monitoring Project, aimed at extending the
number of black hole mass measurements below $\sim 10^7$ M$_\odot$, we
have carried out an imaging campaign in order to generate continuum
light curves for the 13 nearby Seyfert 1 galaxies in our sample. We
used relative aperture photometry to measure the flux variability of
the AGNs compared to several nearby stars. We attempted a simple
host-galaxy starlight subtraction by removing a constant offset from
each flux measurement based on an exponential model of the galaxy
disk. More complete host-galaxy subtraction requires higher resolution
images in order to disentangle the bulge component from the nucleus,
which we will do using future \emph{HST} observations. These AGN
continuum light curves are compared to H$\beta$ emission-line light
curves in Paper III and other optical recombination lines in
\cite{Bentz_2009c} to measure BLR sizes and black hole masses.

There is a wide range of variability in our sample of 13 AGNs.  About
a third of the objects (notably Arp 151, Mrk 1310, NGC 5548, and NGC
6814) exhibit large variations in both the $B$- and $V$-band light
curves throughout most of the campaign. Many of the objects, however,
show a single event or an occasional moderate change in the continuum
level. Other objects (Mrk 142, IC 4218, MCG-06-30-15, and IC 1198)
show very little change in the continuum flux.

We do not find convincing evidence of a measurable time lag between
the $B$- and $V$-band light curves for any of the objects in the
sample. For all but the most variable objects, we found no significant
variations in the $B-V$ color. We attribute the color variability that
is seen in Arp 151, Mrk 1310, and NGC 6814 to incomplete host-galaxy
subtraction.

In addition to the comparison of these AGN continuum light curves to
the H$\beta$ emission-line light curves and the \emph{HST} program to
quantify the amount of host galaxy starlight, future work will compare
the AGN continuum light curves to other emission-line light curves,
such as H$\alpha$, H$\gamma$, and He II in order to determine time
lags and black hole masses. With the measured black hole masses, we
will place the objects in the sample on the
$M_\mathrm{BH}-\sigma_{\star}$ and the
$M_\mathrm{BH}-L_{\mathrm{bulge}}$ relationships. Since MAGNUM has the
capability of simultaneously obtaining optical and near-infrared
images, for Mrk 1310, MCG-06, NGC 5548, and NGC 6814 we have
near-infrared images in addition to the optical images. Future work
will use the near-infrared and optical continuum light curves to learn
more about the size and structure of the dust torus of AGNs. Finally,
future work will investigate whether a structure function analysis can
be used to determine characteristic timescales of variability for
these objects.

\acknowledgements

We thank Mansi Kasliwal for assistance with the P60 scheduling. This
work was supported by National Science Foundation (NSF) grants
AST-0548198 (UC Irvine), AST-0607485 (UC Berkeley), AST-0642621 (UC
Santa Barbara), and AST-0507450 (UC Riverside), as well as by the
TABASGO Foundation (UC Berkeley).  KAIT and its ongoing operation were
made possible by donations from Sun Microsystems, Inc., the
Hewlett-Packard Company, AutoScope Corporation, Lick Observatory, the
NSF, the University of California, the Sylvia \& Jim Katzman
Foundation, and the TABASGO Foundation.  The MAGNUM project has been
supported partly by the Grant-in-Aid of Scientific Research (10041110,
10304014, 12640233, 14047206, 14253001, and 14540223) and COE Research
(07CE2002) of the Ministry of Education, Culture, Sports, Science \&
Technology of Japan. The work of D.S. was carried out at the Jet
Propulsion Laboratory, California Institute of Technology, under a
contract with NASA.  A.V.F. thanks the Aspen Center for Physics, where
he participated in a workshop on Wide-Fast-Deep Surveys while this
paper was nearing completion.  This research has made use of the
NASA/IPAC Extragalactic Database (NED) which is operated by the Jet
Propulsion Laboratory, California Institute of Technology, under
contract with NASA.

\end{document}